\documentclass[final,3p,times]{elsarticle}

\usepackage{amssymb}
\usepackage{graphicx}
\usepackage{epstopdf, epsfig}
\usepackage{placeins}
\usepackage{xcolor}
\usepackage{amsmath}
\usepackage[colorlinks,citecolor=red,urlcolor=blue,bookmarks=false]{hyperref}
\usepackage[capitalise]{cleveref}
\usepackage{flexisym}
\usepackage{booktabs} 
\usepackage{verbatim} 
\usepackage{algorithm}
\usepackage{siunitx}
\usepackage[normalem]{ulem}

\creflabelformat{equation}{#2#1#3}
\crefformat{appendix}{#2#1#3}

\newcommand{\gv}[1]{\ensuremath{\mbox{\boldmath$ #1 $}}} 
\newcommand{\bv}[1]{\ensuremath{\boldsymbol{#1}}} 

\newcommand{\RN}[1]{\textup{\uppercase\expandafter{\romannumeral#1}}}
\newcommand{\order}[1]{\mathcal O \left( #1 \right)}

\newcommand{\Rey}{\ensuremath{Re}}

\newcommand{\Ca}{\ensuremath{Cau}}
\newcommand{\Mn}{\ensuremath{M}}

\newcommand{\CFL}{\ensuremath{\textrm{CFL}}}
\newcommand{\AR}{\ensuremath{AR}}

\definecolor{revred}{HTML}{b00002}
\definecolor{revbloo}{HTML}{5f80d3}
\definecolor{revgreen}{HTML}{54c408}

\newcommand{\blue}[1]{\textcolor{blue}{#1}}

\journal{Journal of Computational Physics}

\begin{document}

\begin{frontmatter}


\title{Soft, slender and active structures in fluids: embedding Cosserat rods in vortex methods}

\author[1]{Arman Tekinalp\fnref{fn1}}
\author[1]{Yashraj Bhosale\fnref{fn1}}
\author[1]{Songyuan Cui}
\author[1]{Fan Kiat Chan}
\author[1,2,3]{Mattia Gazzola\corref{cor1}} \ead{mgazzola@illinois.edu}
\cortext[cor1]{Corresponding author}
\fntext[fn1]{Equal contribution.}
\address[1]{Mechanical Sciences and Engineering, University of Illinois 
Urbana-Champaign, Urbana, IL 61801, USA}
\address[2]{Carl R. Woese Institute for Genomic Biology, University of Illinois 
Urbana-Champaign, Urbana, IL 61801, USA}
\address[3]{National Center for Supercomputing Applications, University of Illinois 
Urbana-Champaign, Urbana, IL 61801, USA}


\author{}

\address{}

\begin{abstract}
We present a hybrid Eulerian-Lagrangian method for the direct simulation of three-dimensional, heterogeneous structures made of soft fibers and immersed in incompressible viscous fluids. Fiber-based organization of matter is pervasive in nature and engineering, from biological architectures made of cilia, hair, muscles or bones to polymers, composite materials or soft robots. In nature, many such structures are adapted to manipulate flows for feeding, locomotion or energy harvesting, through mechanisms that are often not fully understood. While simulations can support the analysis (and subsequent translational engineering) of these systems, extreme fibers’ aspect-ratios, large elastic deformations and two-way coupling with three-dimensional flows, all render the problem numerically challenging. To address this, we couple Cosserat rod theory, which exploits fibers’ slenderness to capture their dynamics in one-dimensional, accurate fashion, with vortex methods via a penalty immersed boundary technique. The favorable properties of the resultant hydroelastic solver are demonstrated against a battery of benchmarks, and further showcased in a range of multi-physics scenarios, involving magnetic actuation, viscous streaming, biomechanics, multi-body interaction, and self-propulsion.
\end{abstract}

\begin{graphicalabstract}
\begin{figure}[!ht]
\centering
\includegraphics[width=\textwidth]{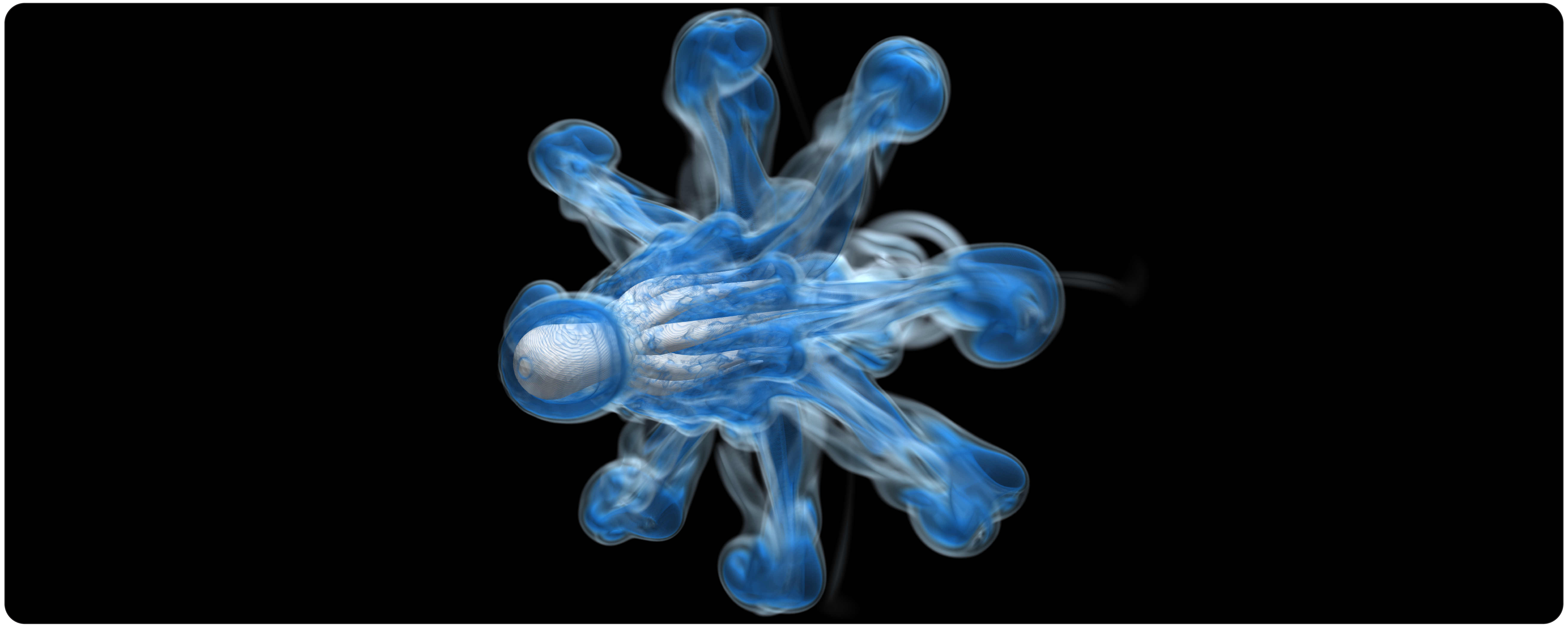}
\label{fig:gabs}
\end{figure}
\end{graphicalabstract}

\begin{highlights}
\item Framework for simulating heterogeneous structures made of slender elastic bodies and immersed in viscous flows.
\item Two-way coupling between Cosserat rod theory (slender elastic bodies) and vortex methods (incompressible viscous flows).
\item Rigorous validation and convergence analysis against a battery of experimental and computational benchmarks.
\item Inclusion of a wide range of multi-physics demonstrations from muscular actuation and bio-locomotion to viscous streaming and magnetism.
\item Scalability enabled via distributed computing for simulating large-scale problems requiring fine spatial resolutions.
\end{highlights}


\end{frontmatter}

\section{Introduction}\label{sec:intro}
This paper presents a hybrid Eulerian-Lagrangian formulation that captures the two-way coupling between multiple heterogeneous, elastic, slender bodies and incompressible viscous fluids. 
Soft slender structures are ubiquitous and pervasive in natural and artificial systems, in active and passive settings, and across scales, from muscles, tendons and bones that make up full organisms \cite{kier2007arrangement,zhang2019modeling,tekinalp2023topology}, to polymers, textiles \cite{prior2014helical,weeger2018nonlinear,arne2010numerical,klar2009hierarchy}, metamaterials \cite{gu2020magnetic,weiner2020mechanics,bhosale2022micromechanical} and soft robots \cite{zhang2019modeling, aydin2019neuromuscular}.
In nature, the interaction between these structures and flow gives rise to a wide array of complex phenomena, ranging from ciliated organisms' feeding strategies~\cite{gilpin2020multiscale}, cephalopods' stereotypical reaching and grasping motions~\cite{flash2005motor,sumbre2001control,tekinalp2023topology}, hagfish's predatory behavior and body knotting~\cite{zintzen2011hagfish}, aquatic animals' locomotion~\cite{Carling:1998,gazzola2012c,verma2018efficient,novati2017synchronisation,bergmann2011modeling,bergmann2014accurate,bergmann2016bioinspired,kazakidi2015vision,huffard2006locomotion}, to the interplay between marine vegetation growth and ocean currents~\cite{mattis2015immersed}.
Simulating and analyzing these systems can be valuable to  understanding and/or engineering them.
However, the presence of extreme aspect ratios in these systems, along with difficulties associated with large elastic deformations and three-dimensional flow coupling, gives rise to significant numerical challenges.

To address these challenges, here we combine Cosserat rod theory and vortex methods. 
Vortex methods~\cite{beale1982vortex,leonard1985computing}, a class of techniques used to resolve flow-structure interaction, solve the velocity-vorticity formulation of momentum equations (as opposed to the velocity--pressure formulation used in other methods), while discretizing solid and fluid phases in Eulerian, Lagrangian, or hybrid fashions.
These methods entail a number of favorable features: guaranteed convergence, accuracy, use of efficient Fourier transforms for solving Poisson equations, compact support of vorticity leading to automatic local ($r$-)adaptivity, natural treatment of far-field boundary conditions, ability to model complex solid morphologies, computational economy and software scalability rivaling traditional velocity-pressure methods \cite{beale1982vortex,leonard1985computing,raviart1985analysis,cottet2000vortex,winckelmans2004vortex,koumoutsakos2005multiscale,coquerelle2008vortex,gazzola2011simulations,gazzola2012c,van2013optimal,gazzola2014scaling,rossinelli2015mrag,bernier2019simulations,rasmussen2011multiresolution}.
These advantages position vortex method as a versatile option to capture the dynamics of unsteady, complex bodies \cite{gazzola2011simulations,gazzola2012c,van2013optimal,bernier2019simulations} across scales \cite{rossinelli2015mrag,Gazzola:2014,Gazzola:2012a}, to deal with contact physics \cite{coquerelle2008vortex}, multiphase \cite{lorieul2018development} and compressible flows \cite{eldredge2002vortex,parmentier2018vortex}, in 2D as well as in 3D \cite{van2013optimal, winckelmans1993contributions,ploumhans2000vortex}.
Despite this versatility, little effort has been made to capitalize on these advantages to resolve two-way coupled dynamics between multiple heterogeneous elastic bodies and the surrounding viscous fluid.

Recently, advancements have been made in this direction~\cite{bhosale2021remeshed}, by integrating inverse map techniques \cite{kamrin2012reference} and vortex method to capture the interaction between bulk rigid and elastic bodies and the surrounding fluid. 
Despite demonstrating versatility, accuracy, and robustness in several multiphysics scenarios, the simulation of fiber-based systems remains hindered by the high spatial resolutions required to resolve geometric scales disparities. 
One possible way to address this challenge is to utilize Cosserat rod theory~\cite{cosserat1909theorie}, which exploits the slenderness of fibers to capture their three-dimensional motions and deformations via a one-dimensional Lagrangian representation. 
This representation entails a number of attractive features, namely, accuracy \cite{gazzola2018forward}, linear scaling of computational complexity with rod resolution \cite{gazzola2018forward}, seamless integration of non-linear constitutive models \cite{zhang2019modeling,tekinalp2023topology} and external environmental effects such as friction, contact \cite{zhang2019modeling,zhang2021friction}, magnetism \cite{yao2023adaptive}, and hydrodynamic loads \cite{ando2021coupled,tschisgale2020immersed}. 
These advantages have resulted in the extensive adoption of elastic rod theory across a wide range of scenarios, including textile manufacturing \cite{prior2014helical,weeger2018nonlinear}, plant mechanics \cite{porat2020general,porat2023mechanical}, artificial muscle modeling \cite{charles2019topology,pawlowski2018dynamic}, musculoskeletal and biological tissue modeling~\cite{zhang2019modeling,pai2005fast,tekinalp2023topology,chang2009modelling}, bio-hybrid machines \cite{zhang2019modeling,pagan2018simulation, aydin2019neuromuscular,wang2021computationally}, 3D printing \cite{weeger2016optimal,perez2015design}, DNA strands \cite{chirikjian2015conformational,welch2020kobra}, meta-materials \cite{kumar2016helical,gupta2019phonons, weiner2020mechanics,bhosale2022micromechanical}, model-based \cite{chang2023energy, wang2022sensory, wang2022control, chang2021controlling} and model-free \cite{naughton2021elastica, shih2023hierarchical} control of soft robots.

Here, the coupling between Cosserat rods and vortex methods is achieved via a mixed Eulerian-Langrangian approach, and the derived solver's accuracy, robustness, versatility and scalability is demonstrated through a range of multiphysics scenarios involving magnetically actuated cilia carpets, viscous streaming, biomechanics, multi-body interaction, and self-propulsion. 
The work is organized as follows: governing equations and modeling techniques are described in \cref{sec:gov} and \cref{sec:method}, respectively; proposed algorithm and numerical discretization are detailed in \cref{sec:num}; rigorous benchmarking and convergence analysis is presented in \cref{sec:bmks}; versatility and robustness of the solver is illustrated through a variety of multifaceted cases in \cref{sec:ills}; finally, concluding remarks are provided in \cref{sec:conc}.

\section{Governing equations}\label{sec:gov}

Here, we provide a description of the governing equations and constitutive laws that dictate the dynamics of multiple rigid bodies and elastic slender structures within a viscous fluid.

\subsection{Governing equations for solids and fluids}\label{sec:governing_eqn_solids_and_fluids}
We start by considering a three-dimensional domain \(\Sigma\) that encompasses an incompressible viscous fluid, as well as rigid and elastic bodies.
As a convention, the subscripts $f$, $r$ and $e$ refer to the fluid, rigid body and the elastic body phases, respectively. 
We use the notations \(\Omega_{e,i}\) and \(\partial\Omega_{e,i}\) (where \(i = 1, \dots, N_e\)) to represent the support and boundaries of the elastic solids, and \(\Omega_{r,j}\) and \(\partial\Omega_{r,j}\) (where \(j = 1, \dots, N_r\)) for the rigid solids. By defining \(\Omega = \Omega_{e,i} \cup \Omega_{r,j}\) as the region occupied by solid material, the remaining region \(\Sigma - \Omega\) is occupied by the fluid. The balance of linear and angular momentum in the fluid domain (for infinitesimal Eulerian volumes \(d\mathbf{x}\)) leads to the incompressible Navier--Stokes equations
\begin{equation}
\label{eqn:ns}
    \bv{\nabla} \cdot \gv{v}_f = 0; ~~~~ \frac{\partial \gv{v}_f}{\partial t} + \left( \gv{v}_f \cdot \bv{\nabla} \right) \gv{v}_f = -\frac{1}{\rho_f}\nabla p_f + \nu_f \bv{\nabla}^2 \gv{v}_f + \gv{f}_f ,~~~~\gv{x}\in\Sigma - \Omega
\end{equation}
where \( t \in \mathbb{R}^+ \) represents time, \(\gv{v}_f : \Sigma \times \mathbb{R}^+ \mapsto \mathbb{R}^3\) represents the velocity field, \(\rho_f\) denotes the fluid density, \(\nu_f\) denotes the fluid's kinematic viscosity, \(p_f : \Sigma \times \mathbb{R}^+ \mapsto \mathbb{R}\) represents the hydrostatic pressure field and \( \gv{f}_f : \Sigma \times \mathbb{R}^+ \mapsto \mathbb{R}^3 \) represents the body force field. The body force field \(\gv{f}_f\) can be further decomposed into two parts \(\gv{f}_f = \gv{f}_{f, v} + \gv{f}_{f, \partial \Omega}\), where \(\gv{f}_{f, v}\) is a conservative volumetric force (like gravitational acceleration), and \(\gv{f}_{f, \partial \Omega}\) is a coupling force used to impose the no-slip condition on the fluid-structure boundary \(\partial \Omega\), as described in \cref{sec:ibm}. 
We make the assumption that all the aforementioned fields are sufficiently smooth both in time and space. 
The interaction between the fluid and elastic solid phases occurs solely through boundary conditions, which enforce velocity continuity (no-slip) and traction forces at all interfaces between the fluid and elastic solid.
\begin{equation}
\label{eqn:elastic_bcs}
    \gv{v} = \gv{v}_f = \gv{v}_{e, i},~~~~
    \bv{\sigma}_f \cdot \gv{n} = \bv{\sigma}_{e, i} \cdot \gv{n} ,~~~~\gv{x}\in
    \partial\Omega_{e,i}
\end{equation}
where $\gv{n}$ denotes the unit outward normal vector at the interface \(\partial \Omega_{e,i}\). 
The interfacial velocities in the fluid and \(i^{\textrm{th}}\) elastic body are represented by \(\mathbf{v}_f\) and \(\mathbf{v}_{e, i}\) respectively. 
Similarly, the interfacial Cauchy stress tensors in the fluid and \(i^{\textrm{th}}\) elastic body are denoted as \(\boldsymbol{\sigma}_f\) and \(\boldsymbol{\sigma}_{e, i}\) respectively. Within the regions \(\Omega_{r, j}\) occupied by rigid solids (\(j=1,\dots,N_r\)), the velocities are kinematically constrained to rigid body modes of pure translation and rotation. Consequently, the interaction between the rigid bodies and the fluid domain is mediated only by the no-slip boundary condition
\begin{equation}
\label{eqn:rigid_bcs}
\gv{v} = \gv{v}_f = \gv{v}_{r, j} = \underbrace{\gv{v}_{cmr, j}}_{\textrm{translation}}
+ \underbrace{\gv{\omega}_{r, j} \times (\gv{x} - \gv{x}_{cmr, j})}_{\textrm{rotation}},
~~~~\gv{x}\in \partial\Omega_{r, j}
\end{equation}
where \( \gv{v}_{r, j} \) is the rigid body velocity field, \( \gv{x}_{cmr, j} \) is the center of mass (COM) position, \( \gv{v}_{cmr, j} \) is the center of mass velocity, and \( \gv{\omega}_{r, j} \) is the angular velocity about the COM of the \(j^{\textrm{\scriptsize th}}\) rigid body.

\subsection{Cosserat rod model for soft, slender heterogeneous structures}\label{sec:cosserat_rod}

To close the above set of equations (\cref{eqn:ns,eqn:elastic_bcs,eqn:rigid_bcs}), it is necessary to specify the governing equations and constitutive laws of the elastic structures. Resolving the dynamics of slender, fiber-like bodies via 3D bulk elasticity frameworks is particularly cumbersome and expensive, given their aspect ratios. We thus adopt the use of Cosserat rod theory. 

\begin{figure}[!ht]
    \centering
\includegraphics[width=\textwidth]{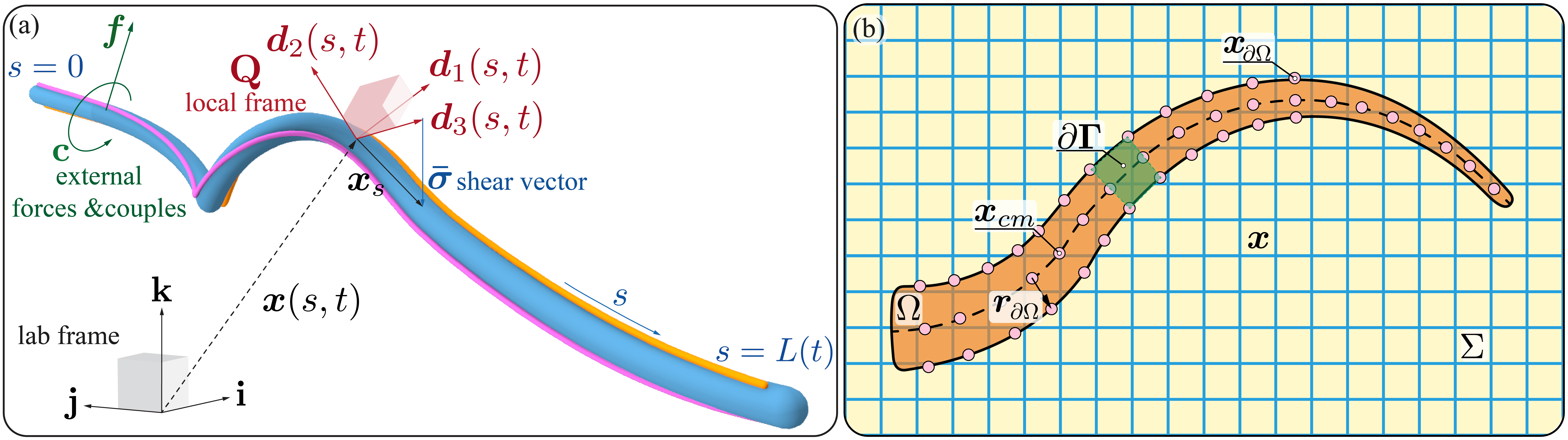}
    \caption{(a) Cosserat rod model. A soft filament (blue) deforming in three-dimensional space is described by a center-line $\gv{x}(s,t)$, and a local orthonormal frame $\bv{Q}(s,t)=\{\gv{d}_{1}, \gv{d}_{2}, \gv{d}_{3}\}$, both functions of centerline's arc-length coordinate $s$ and time $t$. Orange and purple lines around the elastic rod (blue) trace out $\gv{d}_{1}(s,t)$ and $-\gv{d}_{1}(s,t)$, respectively. 
    The shear vector $\gv{\bar{\sigma}}(s,t)$ is quantified by the deviation of  $\gv{d}_{3}(s,t)$ from $\gv{x}_{s}(s,t)$. External force and couple densities are denoted by $\gv{f}(s,t)$ and  $\gv{c}(s,t)$, respectively.
    (b) 
    Penalty immersed boundary (pIBM) setup. An elastic body/rod (orange region) defined on a Lagrangian domain $\Omega$ is immersed in a viscous fluid represented on a fixed Eulerian domain $\Sigma$. 
    The elastic rod/body is discretized into Lagrangian elements along the centerline, separated by nodes $\gv{x}_{cm}(s,t)$ on the same centerline.
    An immersed boundary $\partial \Omega$ body-fitted to the elastic body and discretized via forcing points (pink circles), runs parallel to the Lagrangian elements with $\gv{x}_{\partial \Omega} := \gv{r}_{\partial \Omega} + \gv{x}_{cm} $, where $\gv{r}_{\partial \Omega}$ is the moment arm of the forcing points.
    The green area $\partial \gv{\Gamma}$ refers to the boundary or interface that separates a Lagrangian element from the viscous fluid that surrounds it.
    }
    \label{fig:cosserat_rod_flow_setup}
\end{figure}

Cosserat rod theory captures three-dimensional motions and deformations of slender elastic structures via a one-dimensional Lagrangian representation. 
\Cref{fig:cosserat_rod_flow_setup}a presents the setup of the Cosserat rod theory model. 
Each point on the Cosserat rod is characterized by a center-line $\gv{x}(s,t) \in \mathbb{R}^{3}$ and an orthonormal frame $\bv{Q}(s,t) \in \mathbb{R}^{3 \times 3} = \{\gv{d}_{1}, \gv{d}_{2}, \gv{d}_{3}\}$ (triad of unit vectors $\gv{d}_i \in \mathbb{R}^{3}$), which defines the frame transformations between the laboratory and local frames. 
Here, $s \in [0, L(t)]$ is the center-line arc-length coordinate, where $L$ is the rod's current length, and $t$ is the time. 
Any vector $\gv{v} \in \mathbb{R}^{3}$ defined in the laboratory frame can be transformed into its local frame counterpart $\bar{\gv{v}} \in \mathbb{R}^{3}$, by $\gv{\bar{v}} = \bv{Q}\gv{v}$ and from local to laboratory frame by $\gv{v} = \bv{Q}^{T}\gv{\bar{v}}$. 
For an unshearable and inextensible rod, $\gv{d}_{3}$ is parallel to the rod tangent $\gv{x}_{s} \in \mathbb{R}^{3} = \partial_s \gv{x}$, with $\gv{d}_{1}$ (normal) and $\gv{d}_{2}$ (binormal) spanning the cross-section of the rod. 
Under shear and extension, the rod tangent $\gv{x}_{s}$ and $\gv{d}_{3}$ are no longer parallel and their shift can be quantified via the shear vector $\gv{\bar{\sigma}} \in \mathbb{R}^{3} = \bv{Q} \left(\gv{x}_{s} - \gv{d}_{3} \right)$ in the local frame.
The curvature vector $\gv{\kappa}(s,t) \in \mathbb{R}^{3}$ encodes $\bv{Q}$'s rotations along the rod through the relation $\partial_s \mathbf{d}_{j} = \gv{\kappa} \times \gv{d}_{j}$.
The rate of change of $\bv{Q}$ is given by $\partial_t \gv{d}_{j} = \gv{\omega} \times \gv{d}_{j}$, where $\gv{\omega} \in \mathbb{R}^{3}$ corresponds to the angular velocity. 
After defining the velocity of the centerline $\gv{v} \in \mathbb{R}^{3} = \partial_t \gv{x}$, second area moment of inertia $\bv{\bar{I}} \in \mathbb{R}^{3 \times 3}$, cross-sectional area $A$, and density $\rho$, the dynamics \cite{Gazzola:2018} of a soft slender body is described by
\begin{equation} 
    \partial_{t}^{2} \left( \rho A \gv{x} \right)= \partial_s \left(\bv{Q}^{T} \gv{\bar{n}}\right) + \gv{f}
    \label{eqn:linear_momentum}
\end{equation}
\begin{equation}\label{eqn:angular_momentum}
    \partial_t \left( \rho \bv{\bar{I}} \gv{\bar{\omega}} \right) = \partial_s \gv{\bar{\tau}} + \gv{\bar{\kappa}} \times \gv{\bar{\tau}} + \left(\bv{Q} \gv{x}_s \times \gv{\bar{n}}\right)
     + \left(\rho \bv{\bar{I}} \gv{\bar{\omega}}\right) \times \gv{\bar{\omega}} + \bv{Q} \gv{c}
\end{equation}
where \cref{eqn:linear_momentum,eqn:angular_momentum} represent linear and angular momentum balance at every cross section, $\gv{\bar{n}}$ and $\gv{\bar{\tau}}$ are the internal forces and torques, $\gv{f}$ and $\gv{c}$ are the external force and torque densities, respectively. 
To close the above \cref{eqn:linear_momentum,eqn:angular_momentum}, it is necessary to specify the form of the internal forces  ($\gv{\bar{n}}$) and torques ($\gv{\bar{\tau}}$) generated in response to shear ($\gv{\bar{\sigma}}$) and bend/twist ($\gv{\bar{\kappa}}$) strains. 
In this study, for simplicity, we assume a perfectly elastic material so that the stress–strain relations are linear. 
We note that Cosserat rod theory is not limited to modeling linear elasticity, but can be extended to non-linear elastic materials \cite{zhang2019modeling,tekinalp2023topology}.
Within the linear elasticity model, the internal forces $\gv{\bar{n}}$ can be expressed as
\begin{equation}
    \gv{\bar{n}} = \bv{\bar{S}} \left(\gv{\bar{\sigma}} - \gv{\bar{\sigma}}_0 \right)
    \label{eqn:force_contitutive}
\end{equation}
where $\gv{\bar{\sigma}}_0$ is the intrinsic shear strain and $\bv{\bar{S}} \in \mathbb{R}^{3 \times 3} = \text{diag}\left(\bar{S}_1, \bar{S}_2, \bar{S}_3 \right) = \text{diag}\left(\alpha_c G A, \alpha_c G A, E A \right)$ is the shear/stretch stiffness matrix. 
Here, $\bar{S}_{1}$, $\bar{S}_{2}$ and $\bar{S}_{3}$ are the shear/stretch rigidities about $\gv{d}_1$, $\gv{d}_2$ and $\gv{d}_{3}$, respectively. 
Additionally, $E$, $G$ and $\alpha_c$ correspond to the Young's modulus, the shear modulus, and the shear correction factor, respectively (for details, see \cite{Gazzola:2018}). 
Similarly, the internal torques $\gv{\bar{\tau}}$ can be expressed as
\begin{equation}
    \gv{\bar{\tau}} = \bv{\bar{B}} \left(\gv{\bar{\kappa}} - \gv{\bar{\kappa}}_0 \right)
    \label{eqn:torque_contitutive}
\end{equation}
where $\gv{\bar{\kappa}}_0$ is the intrinsic rod curvature and $\bv{\bar{B}} \in \mathbb{R}^{3 \times 3} = \text{diag}\left(\bar{B}_1, \bar{B}_2, \bar{B}_3 \right) = \text{diag}\left(E \bar{I}_1, E \bar{I}_2, G \bar{I}_3 \right)$ is the bend/twist stiffness matrix.
Here, $\bar{B}_{1}$, $\bar{B}_{2}$ and $\bar{B}_{3}$ are the bend/twist rigidities about $\gv{d}_1$, $\gv{d}_2$ and $\gv{d}_{3}$, respectively. 
Additionally, $\bar{I}_{1}$, $\bar{I}_{2}$ and $\bar{I}_{3}$ correspond to the values of second moment of inertia about $\gv{d}_1$, $\gv{d}_2$ and $\gv{d}_{3}$, respectively. 

The Cosserat rod representation presented above enables a number of favorable features:
(1) it captures 3D dynamic effects and all modes of deformation (bending, twist, shear, stretch), particularly relevant in the case of biological or elastomeric materials \cite{Gazzola:2018,tekinalp2023topology,porat2023mechanical};
(2) non-linear constitutive models can be seamlessly integrated via \cref{eqn:force_contitutive,eqn:torque_contitutive} \cite{zhang2019modeling,tekinalp2023topology,porat2023mechanical};
(3) complexity scales as $O(n)$ with axial resolution \cite{Gazzola:2018}; 
(4) kinematic or dynamic boundary conditions can be directly enforced, allowing the assembly of multiple rods into complex architectures \cite{zhang2019modeling,tekinalp2023topology}; (5) continuum and/or distributed actuation and environmental loads can be directly incorporated via $\gv{f}$ and $\gv{c}$ into the Cosserat rod dynamics, rendering the inclusion of contact dynamics \cite{weiner2020mechanics,bhosale2022micromechanical}, friction \cite{weiner2020mechanics,bhosale2022micromechanical}, muscular activity \cite{zhang2019modeling,tekinalp2023topology}, magnetism \cite{magnetopyelastica2023} or hydrodynamics \cite{ando2021coupled,tschisgale2020immersed} straightforward.

\section{Methodology}\label{sec:method}

Having established the fundamental governing equations and boundary conditions, we will now introduce the methodologies employed for solving these coupled equations.
We employ the Cosserat rod equations to track the dynamics and deformations of elastic slender bodies, and the immersed boundary method to solve the coupling problem in a hybrid Eulerian/Lagrangian vortex method framework.

\subsection{Vortex methods}\label{sec:rvm}
\noindent
Vortex methods consider the vorticity form of the Navier--Stokes equations, obtained by taking the curl of \cref{eqn:ns}
\begin{equation}
\label{eqn:vort_ns}
    \frac{\partial \gv{\omega}_f}{\partial t} + \left( \gv{v}_f \cdot \bv{\nabla} \right) \gv{\omega}_f - \left( \gv{\omega}_f \cdot \bv{\nabla} \right) \gv{v}_f = \nu_f \bv{\nabla}^2 \gv{\omega}_f + \bv{\nabla} \times \gv{f}_f,
\end{equation}
where \(\gv{\omega}_f : \Sigma \times \mathbb{R}^+ \mapsto \mathbb{R}^3 := {\nabla} \times \gv{v}_f\) represents the vorticity field in the fluid. With pressure $p$ eliminated from the governing equations, an incompressible velocity
field is then directly recovered from the vorticity by solving a Poisson equation using
appropriate boundary conditions on \( \Sigma \)
\begin{equation}
\label{eqn:poisson}
    \gv{\nabla}^2 \gv{\psi}_f = - \gv{\omega}_f; ~~~~
    \gv{v}_f = \gv{\nabla} \times \gv{\psi}_f
\end{equation}
where $\gv{\psi}_f : \Sigma \times \mathbb{R}^+ \mapsto \mathbb{R}^3$ (in 2D $\gv{\psi}_f$ corresponds to the streamfunction). 
Favorable features of this approach are: guaranteed convergence, accuracy, use of efficient Fourier transforms for solving Poisson equations, compact support of vorticity leading to automatic local ($r$-)adaptivity, natural treatment of far field boundary conditions, ability to model complex solid morphologies, computational economy and software scalability \cite{beale1982vortex,leonard1985computing,raviart1985analysis,cottet2000vortex,winckelmans2004vortex,koumoutsakos2005multiscale,coquerelle2008vortex,gazzola2011simulations,gazzola2012c,van2013optimal,gazzola2014scaling,rossinelli2015mrag,bernier2019simulations,rasmussen2011multiresolution}.

\subsection{Immersed boundary method}\label{sec:ibm}

We connect the forcing terms $\gv{f}$ from the Navier--Stokes equations (\cref{eqn:vort_ns}) and the Cosserat rod equations (\cref{eqn:linear_momentum,eqn:angular_momentum}) via the immersed boundary method.
Immersed boundary methods (IBMs) are widely adopted computational techniques for simulating fluid-structure interaction problems involving complex geometries \cite{peskin1972flow,Huang:2019}.
In IBMs, the fluid domain is typically discretized using an Eulerian grid, while the immersed object is represented on a Lagrangian grid that can move freely with respect to the underlying Eulerian fluid mesh. The interaction between the fluid and solid phases is modeled using a force distribution function, applied on the Lagrangian and Eulerian grids, to achieve two-way fluid-structure interactions. 
Due to their versatility, IBMs have been employed in a range applications, from microfluidics \cite{han2018spontaneous,jayathilake2012three} and biomedical engineering \cite{yuan2014ib,peskin1972flow} to the aerospace industry \cite{de2007immersed}.

In this study, we adopt the penalty immersed boundary method (pIBM), a flavor of IBMs that adds penalty forces to the governing equations of both fluid and solid phases, to enforce appropriate boundary conditions (\cref{eqn:rigid_bcs,eqn:elastic_bcs}) at the fluid-solid interface \cite{goldstein1993modeling,kim2007penalty}. 
\Cref{fig:cosserat_rod_flow_setup}b presents the setup of the penalty immersed boundary method---an elastic body $(\Omega)$ fitted with an immersed boundary at the fluid-solid interface $(\partial \Omega)$ is submerged in a viscous fluid $\Sigma$. 
The steps involved in the pIBM formulation for the two-way coupling can be briefly summarized as follows: 
(1) interpolate the flow velocity $\gv{v}_{f}$ from the Eulerian grid $\left(\Sigma\right)$ and the body velocity $\gv{v}_{i}$ from the Lagrangian domain $\left(\Omega\right)$ onto the immersed boundary $\left(\partial \Omega \right)$; 
(2) compute the coupling force between the two phases based on the deviation between the interpolated velocities;
(3) transfer the coupling forces via the forcing terms $\gv{f}$ to the fluid ($\Sigma$) and the solid ($\Omega$) domains, respectively.

We next present a detailed step-by-step implementation of the pIBM formulation. 
We begin by computing the velocity of the fluid phase on the Lagrangian boundary $\partial \Omega$ at position $\gv{x}_{\partial \Omega}$, via interpolation
\begin{equation}
        \gv{v}_{f,\partial \Omega} = \int_{\Sigma} \gv{v}_{f}(\gv{x})~\delta \left(\gv{x}_{\partial \Omega} - \gv{x}\right) d\gv{x}
        \label{eqn:ibm_flow_vel_interpolation}
\end{equation}
where $\gv{v}_{f}(\gv{x})$, $\gv{v}_{f,\partial \Omega}$ correspond to the fluid velocity at the Eulerian position $\gv{x}$ and on the immersed boundary ($\partial \Omega$), and $\delta(\cdot)$ is the Dirac delta function.
Simultaneously, we compute the velocity of the \(i^{\textrm{\scriptsize th}}\) body on $\partial \Omega$ ($\gv{v}_{i,\partial \Omega}$) as
\begin{equation}
    \gv{v}_{i,\partial \Omega} = \gv{v}_{i,cm} + \gv{\omega}_{i,cm} \times \gv{r}_{i,\partial \Omega}
\end{equation}
where $\gv{v}_{i, cm}$, $\gv{\omega}_{i, cm}$, and  $\gv{r}_{i,\partial \Omega}$ represent the translational and rotational velocity of the center of mass (COM), and the moment arm of the immersed boundary from COM, respectively. With the velocities of both the phases on $\partial \Omega$ known, the interaction force $\gv{f}_{\partial \Omega}$ between the fluid and the \(i^{\textrm{\scriptsize th}}\) body is then computed as
\begin{equation}
    \gv{f}_{\partial \Omega} = - \int_0^t \alpha \left(\gv{v}_{i,\partial \Omega} - \gv{v}_{f,\partial \Omega}\right) dt - \beta \left( \gv{v}_{i,\partial \Omega} - \gv{v}_{f,\partial \Omega} \right)
\end{equation}
where $\alpha$ and $\beta$ are the coupling stiffness constant and the coupling damping constant, respectively. 
To enforce the fluid-solid interfacial conditions (\cref{eqn:rigid_bcs,eqn:elastic_bcs}) accurately, we require $\alpha \gg 1$ (although too large values often lead to numerical instabilities \cite{goldstein1993modeling}). 
Typically, $\alpha=\mathcal{O}(10^4)$ and $\beta=\mathcal{O}(10)$ are observed to strike a balance between accuracy and stability \cite{huang2007simulation,tian2014fluid}.
The final step in the pIBM formulation involves transferring equal and opposite coupling forces to the fluid and solid phases. 
The coupling force on the fluid $\gv{f}_{f,\partial \Omega}$ at the Eulerian position $\gv{x}$ is computed as
\begin{equation}
    \gv{f}_{f, \partial \Omega}(\gv{x}) = -\int_{\partial \Omega} \gv{f}_{\partial \Omega} (\gv{x}_{\partial \Omega})~\delta \left(\gv{x} - \gv{x}_{\partial \Omega} \right) d\gv{x}_{\partial \Omega}.
\end{equation}
Before computing the coupling forces on the elastic body, note that the elastic rod/body is discretized into Lagrangian elements along the centerline (green zone, \cref{fig:cosserat_rod_flow_setup}b), defined by the nodes $\gv{x}_{cm}(s,t)$ on the same centerline.
Then the coupling forces $\gv{f}_{i, \partial \Omega}$ and torques $\gv{c}_{i, \partial \Omega}$ acting  on a single Lagrangian element of the \(i^{\textrm{\scriptsize th}}\) body $\Omega_{i}$  are computed as
\begin{equation}
    \gv{f}_{i, \partial \Omega} = \int_{\partial \Gamma} \gv{f}_{\partial \Omega}~d\gv{x}_{\partial \Omega}~~~~\gv{c}_{i, \partial \Omega} = \int_{\partial \Gamma} \left(\gv{r}_{i,\partial \Omega} \times \gv{f}_{\partial \Omega} \right) ~d\gv{x}_{\partial \Omega}
\end{equation}
where $\partial \Gamma$ refers to the fluid-solid interface of the Lagrangian element.
The pIBM formulation presented above is
(1) relatively easy to implement and can be integrated into existing numerical codes  \cite{kim2016penalty};
(2) decouples the representation of the fluid and solid phases, simplifying mesh generation (no conforming meshes \cite{mittal2005immersed}); 
(3) reduces computational cost by avoiding remeshing in the case of moving boundaries \cite{Huang:2019};
(4) enables efficiency and parallelization performance typical of structured grids  \cite{tian2014fluid};
(5) accommodates different phase representations (fluid or solid), enabling relatively straightforward integration of independently developed methodologies for fluid, solid, or other multi-physics components \cite{kim2008numerical,huang2018improved}.

In the next section, we proceed to describe the numerical discretization of the elements described above, with a detailed step-by-step explanation of our algorithm.

\section{Numerical discretization and algorithm}\label{sec:num}

We begin by spatially discretizing the system of equations for the fluid phase (\cref{eqn:vort_ns,eqn:poisson}) via an Eulerian Cartesian grid of uniform spacing \(h\) to create our computational domain \(\Sigma_{h}\). The continuous flow fields previously defined are now replaced with their discrete counterparts on \(\Sigma_{h}\).
For the elastic fibers, the Cosserat rod equations (\cref{eqn:linear_momentum,eqn:angular_momentum}) are discretized over a Lagrangian grid, consisting of discrete elements along the centerline with uniform spacing $d s$. 
Finally, the fluid-solid interfaces (immersed boundaries) are discretized using Lagrangian grids $\partial \Omega_h$ body-fitted to the solid domain $\Omega_h$, with uniform grid spacing $h_{\partial \Omega}$.
The spacing between Lagrangian grid points $h_{\partial \Omega}$ is calculated based on the spacing of Eulerian grid points $h$, ensuring that each Lagrangian point spans across 1-2 neighboring Eulerian points.
The implementation of a relative grid spacing policy guarantees the synchronized convergence of fluid-solid dynamics, while also preventing artificial gaps between the two interfaces.

Next, we outline Algorithm \ref{alg} which describes one complete time step, spanning from \(t^n\) to \(t^{n+1}\), assuming that all relevant quantities are already known up to time \(t^n\).

\subsection{Poisson solve and velocity recovery}\label{sec:poisson}

First, we solve the Poisson equation (\cref{poisson_solve}) for the variable \(\gv{\psi}\). Depending on the problem setup, we employ periodic or unbounded boundary conditions. To solve \cref{poisson_solve} on the grid, we use a Fourier-series based solver that operates with \( \order{n \log(n)} \) complexity. This solver takes advantage of the diagonal nature of the Poisson operator in the case of periodic boundaries \cite{hockneycomputer}, allowing for spectral accuracy. For unbounded conditions, we utilize the zero padding technique described by Hockney and Eastwood \cite{hockneycomputer}. Once \(\gv{\psi}\) is obtained on the grid, we calculate the velocity according to \cref{vel_from_psi} using the discrete second-order centered finite difference curl operator.

\begin{algorithm}
\caption{General algorithm}
\label{alg}
\begin{align}
    \textrm{Poisson solve (\cref{sec:poisson})} ~~~ &\bv{\nabla}^2 \gv{\psi}_f^n = - \gv{\omega}_f^n
    \label{poisson_solve} \\
    \textrm{Velocity recovery (\cref{sec:poisson})} ~~~ &\gv{v}_f^n = \bv{\nabla} \times \gv{\psi}_f^n + \gv{V}_{f, \infty}^n
    \label{vel_from_psi} \\
    \textrm{Compute flow velocity on fluid-solid interface (\cref{sec:ibm_computation})} ~~~ &\gv{v}^n_{f,\partial \Omega} = \int_{\Sigma} \gv{v}^n_{f}(\gv{x})~\delta \left(\gv{x}^n_{\partial \Omega} - \gv{x}\right) d\gv{x}
    \label{flow_velocity_on_ibm} \\
    \textrm{Compute body velocity on fluid-solid interface (\cref{sec:ibm_computation})} ~~~ &\gv{v}^n_{i,\partial \Omega} = \gv{v}^n_{i,cm} + \gv{\omega}^n_{i,cm} \times \gv{r}^n_{i,\partial \Omega}
    \label{body_velocity_on_ibm} \\
    \textrm{Compute coupling force on fluid-solid interface (\cref{sec:ibm_computation})} ~~~ &\gv{f}^n_{\partial \Omega} = - \int_0^{t^n} \alpha \left(\gv{v}^n_{i,\partial \Omega} - \gv{v}^n_{f,\partial \Omega}\right) dt - \beta \left( \gv{v}^n_{i,\partial \Omega} - \gv{v}^n_{f,\partial \Omega} \right)
    \label{interaction_force_on_ibm} \\
    \textrm{Transfer coupling force to fluid (\cref{sec:ibm_computation})} ~~~ &\gv{f}^n_{f, \partial \Omega}(\gv{x}) = -\int_{\partial \Omega} \gv{f}^n_{\partial \Omega} (\gv{x}^n_{\partial \Omega})~\delta \left(\gv{x} - \gv{x}^n_{\partial \Omega} \right) d\gv{x}_{\partial \Omega}
    \label{transfer_interaction_force_to_fluid} \\
    \textrm{Transfer coupling force to body (\cref{sec:ibm_computation})} ~~~ &\gv{f}^n_{i, \partial \Omega} = \int_{\partial \Gamma} \gv{f}^n_{\partial \Omega}~d\gv{x}_{\partial \Omega};~~ \gv{c}^n_{i, \partial \Omega} = \int_{\partial \Gamma} \left(\gv{r}^n_{i,\partial \Omega} \times \gv{f}^n_{\partial \Omega}\right) ~d\gv{x}_{\partial \Omega}
    \label{transfer_interaction_force_to_body} \\
    \textrm{Fluid (vorticity) update (\cref{sec:vort_update})} ~~~ &\frac{\partial \gv{\omega}_f^n}{\partial t} +
    \left( \gv{v}_f^n \cdot \bv{\nabla} \right) \gv{\omega}_f^n - \left( \gv{\omega}_f^n \cdot \bv{\nabla} \right) \gv{v}_f^n = \bv{\nabla} \times \left ( \gv{f}_{f, v}^n + \gv{f}_{f, \partial \Omega}^n \right ) + \nu_f \bv{\nabla}^2 \gv{\omega}_f^n
    \label{vort_update} \\
    \textrm{Vorticity propagation to next time step (\cref{sec:vort_update})} ~~~ &\gv{\omega}_f^{n + 1} = \gv{\omega}_f^{n}
    \label{vort_reset} \\
    \textrm{Elastic body update (\cref{sec:body_update})} ~~~ &\gv{x}^{n}_{e, i},~\gv{v}^{n}_{e, i},~\bv{Q}^{n}_{e, i},~\gv{\omega}^{n}_{e, i} \xrightarrow{\gv{f}^n_{e, i, \partial \Omega},~ \gv{c}^n_{e, i, \partial \Omega}} \gv{x}^{n+1}_{e, i},~\gv{v}^{n+1}_{e, i},~\bv{Q}^{n+1}_{e, i},~\gv{\omega}^{n+1}_{e, i}
    \label{elastic_body_update} \\
    \textrm{Rigid body update (\cref{sec:body_update})} ~~~ &\gv{x}^{n}_{cmr, j},~\gv{v}^{n}_{cmr, j},~\bv{Q}^{n}_{r, j},~\gv{\omega}^{n}_{r, j} \xrightarrow{\gv{f}^n_{r, j, \partial \Omega},~ \gv{c}^n_{r, j, \partial \Omega}} \gv{x}^{n+1}_{cmr, j},~\gv{v}^{n+1}_{cmr, j},~\bv{Q}^{n}_{r, j},~\gv{\omega}^{n+1}_{r, j}
    \label{rigid_body_update}
\end{align}
\end{algorithm}

\subsection{Immersed boundary computation}\label{sec:ibm_computation}

For each solid body, the velocity of the fluid phase on the Lagrangian boundary $\partial \Omega_h$ at position $\gv{x}_{\partial \Omega_h}$ is computed via the discrete form of \cref{flow_velocity_on_ibm}, which reads
\begin{equation}
    \gv{v}^n_{f,\partial \Omega} = \sum_{\gv{x}_h \in g_h} \gv{v}^n_{f}(\gv{x}_h)~\delta_h \left(\gv{x}^n_{\partial \Omega_h} - \gv{x}_h\right) h^d
\end{equation}
where $d$, $\delta_h(\cdot)$ and $g_h$ correspond to the grid dimension (2 for 2D and 3 for 3D), mollified Delta function, and support of the Delta function, respectively.
The mollified Delta function $\delta_h(\cdot)$ reads 
\begin{equation}
    \delta_h(\gv{x}) = \frac{1}{h^d} \prod_{i=1}^d  W \left(\frac{x_i}{h}\right)
\end{equation}
where $i$ symbolizes the grid index and $W(\cdot)$ is an interpolation kernel. 
In this study, $W(\cdot)$ is the four point Delta function introduced by \citet{peskin2002immersed} and widely used in IBM studies. 
Following interpolation, the velocity of the solid body is then computed on the Lagrangian boundary $\partial \Omega_h$ via \cref{body_velocity_on_ibm}. 
With the velocities of both the phases on $\partial \Omega_h$ known, the interaction force $\gv{f}_{\partial \Omega_h}$ is then computed via \cref{interaction_force_on_ibm}, with the time integral on the RHS of \cref{interaction_force_on_ibm} computed discretely via a left Riemann sum.
Following interaction force computations (\cref{interaction_force_on_ibm}), the coupling forces are transferred to the fluid phase via the discrete form of \cref{transfer_interaction_force_to_fluid}
\begin{equation}
    \gv{f}_{f, \partial \Omega_h}(\gv{x}_h) = -\sum_{\gv{x}_{\partial \Omega_h} \in g_h} \gv{f}_{\partial \Omega_h} (\gv{x}_{\partial \Omega_h})~\delta_h \left(\gv{x}_h - \gv{x}_{\partial \Omega_h} \right) h^{d-1}_{\partial \Omega}
\end{equation}
Simultaneously, equal and opposite forces and corresponding torques are transferred to the \(i^{\textrm{\scriptsize th}}\) body $\Omega_{i}$ via the discrete version of \cref{transfer_interaction_force_to_body}
\begin{equation}
    \gv{f}_{i, \partial \Omega_h} = \sum_{\partial \Gamma_h} \gv{f}_{\partial \Omega_h} h^{d-1}_{\partial \Omega}~~~~\gv{c}_{i, \partial \Omega_h} = \sum_{\partial \Gamma_h} \left(\gv{r}_{i,\partial \Omega_h} \times \gv{f}_{\partial \Omega_h} \right) h^{d-1}_{\partial \Omega}
\end{equation}

\subsection{Fluid phase (vorticity) update}\label{sec:vort_update}

After the transfer of the coupling forces to the fluid (\cref{transfer_interaction_force_to_fluid}), we next update the fluid state (vorticity) via \cref{vort_update}.
The numerics of this step entails computing three main components that modify the vorticity, namely, the body forces $\bv{\nabla} \times \left ( \gv{f}_{f, v}^n + \gv{f}_{f, \partial \Omega}^n \right)$, the diffusion flux $\nu_f \bv{\nabla}^2 \gv{\omega}_f^n$ and the flow inertia flux, that includes the advection $\left( \gv{v}_f^n \cdot \bv{\nabla} \right) \gv{\omega}_f^n$ and stretching $\left( \gv{\omega}_f^n \cdot \bv{\nabla} \right) \gv{v}_f^n$ terms. 
For the body forcing and diffusion flux terms, we compute $\bv{\nabla} \times \left ( \gv{f}_{f, v}^n + \gv{f}_{f, \partial \Omega}^n \right)$ and $\nu_f \bv{\nabla}^2 \gv{\omega}_f^n$ by replacing the curl operator $\bv{\nabla} \times$ and the Laplacian operator $\bv{\nabla}^2$ with their discrete second-order centered finite difference counterparts. 

For the flow inertia terms, we use slightly different formulations in two-dimensional and three-dimensional settings, for the following reasons. 
In 2D flows, where the vortex stretching term $\left( \gv{\omega}_f^n \cdot \bv{\nabla} \right) \gv{v}_f^n$ identically vanishes, the flow inertia term reduces to the advection term $\left( \gv{v}_f^n \cdot \bv{\nabla} \right) \gv{\omega}_f^n$, which is discretized using an ENO3 stencil \cite{liu1994weighted}. 
However, in 3D flows where the vortex stretching term is typically non-vanishing, the same approach cannot be applied in a numerically stable fashion.
The rationale behind this is that since the vorticity $\gv{\omega}_f$ is defined as the curl of a smooth differentiable velocity field $\gv{v}_f$ ($\gv{\omega}_f = \bv{\nabla} \times \gv{v}_f$), theoretically it needs to satisfy the zero divergence constraint ($\bv{\nabla} \cdot \gv{\omega}_f = 0$). While this is trivially satisfied in 2D, in 3D the numerical errors resulting from the advection and stretching terms create spurious divergence in the vorticity field, leading to long-time numerical instabilities.
To mitigate this issue, based on insights from pseudospectral methods \cite{mortensen2016massively}, we solve \cref{vort_update} via two back-to-back steps
\begin{equation}
    \frac{\partial \gv{\omega}_f^n}{\partial t} + \bv{\nabla} \times \left(\gv{\omega}_f^n \times \gv{v}_f^n \right) = \bv{\nabla} \times \left ( \gv{f}_{f, v}^n + \gv{f}_{f, \partial \Omega}^n \right ) + \nu_f \bv{\nabla}^2 \gv{\omega}_f^n
    \label{eqn:rotational_ns}
\end{equation}
\begin{equation}
     \gv{\omega}_f^n = L^p(\gv{\omega}_f^n) 
    \label{eqn:vort_filter}
\end{equation}
where the first step (\cref{eqn:rotational_ns}) involves reformulating the flow inertia terms into a rotational form $\bv{\nabla} \times \left(\gv{\omega}_f^n \times \gv{v}_f^n \right)$ (curl of a vector field). 
The second step (\cref{eqn:vort_filter}), similar to dealiasing in pseudospectral methods,  corresponds to applying a compact discrete (only using neighbouring grid values) filter $L^p(\cdot)$ on the vorticity outputted from \cref{eqn:rotational_ns}. 
The superscript $p$, taking integer values, corresponds to the filter order and qualitatively maps to the strength of the filter; lower values of $p$ imply stronger filtering action. 
We set $p = 5$ for all 3D simulations to avoid artificial dampening of vorticity while preserving stability of our simulations. 
Additional details on the numerical implementation and transfer function of the compact discrete filters can be found in \citet{jeanmart2007investigation}.

Following the evaluation of all forcing terms in \cref{vort_update}, we then perform temporal integration using a forward Euler timestep. 
The vorticity after this step is then carried over to the next timestep (\cref{vort_reset}).

\subsection{Elastic and rigid body update}\label{sec:body_update}

In \cref{elastic_body_update} we resolve the elastic rod dynamics. We discretize the Cosserat rod equations (\cref{eqn:linear_momentum,eqn:angular_momentum}) on a Lagrangian grid of uniform spacing $d s$ along the centerline (thus forming a set of discrete cylindrical elements of length $d s$). 
For terms on the right-hand side of \cref{eqn:linear_momentum,eqn:angular_momentum}  involving spatial differentiation along the centerline, we use second-order centered finite differences to compute the derivatives.
Finally, with the internal and external, forces and torques known, we evolve the kinematic state (positions, orientations and velocities) of all the elastic rods (\cref{elastic_body_update})
using the symplectic, second-order position Verlet time integration scheme. Similarly, rigid bodies are updated in \cref{rigid_body_update}. Numerical details and validation of the Cosserat rod dynamics solver can be found in \cite{gazzola2018forward}, while the implementation can be found in the open-source software \texttt{PyElastica} \cite{pyelastica2023}.

\subsection{Restrictions on simulation time step}\label{sec:timestep}

In our algorithm, we face three main restrictions on the time step size due to the existence of distinct time scales in the coupling problem.
The first restriction is a typical CFL (Courant--Friedrich--Lewy) condition, that stems from the advection and stretching of vorticity inside the fluid
\begin{equation}
    \label{dt_CFL}
    \Delta t_1 \leq h ~ \CFL ~ ||\gv{v}_{f}||_{\infty}^{-1}
\end{equation}
where $\gv{v}_f$ refers to the velocity field inside the fluid. The second restriction is the Fourier condition that ensures stability with regards to explicit time discretization
of the viscous stresses inside the fluid
\begin{equation}
    \label{dt_diffusion}
    \Delta t_2 \leq k ~ h^2 / (2d \nu_{f})
\end{equation}
where $k$ is a constant usually set to be \( \leq 1\). Here we set $k = 0.9$  throughout.
Finally, the third restriction stems from the need to resolve shear waves inside elastic solids.
This is a CFL-like restriction dependent on the shear wave speed $c_{sh}$
\begin{equation}
    \label{dt_shear}
    \Delta t_3 \leq  ds ~ \CFL  ~ c_{sh}^{-1} = ds ~\CFL  ~ \sqrt{\rho_e / G}.
\end{equation}
Here, $ds$, $\rho_e$ and $G$ correspond to the rod's Lagrangian grid spacing, solid density and shear modulus, respectively. 
Combining \cref{dt_CFL,dt_diffusion,dt_shear}, we obtain the final criterion to adapt the time
step during simulation
\begin{equation}
    \Delta t = \min{\left[ \Delta t_1, \Delta t_2, \Delta t_3 \right] }.
\end{equation}
We observe (a posteriori) that in several cases investigated in \cref{sec:bmks} and \cref{sec:ills}, the time-step restriction from the high elastic shear wave speeds $\Delta t_3$, rather than the CFL condition $\Delta t_1$ or the diffusion limit $\Delta t_2$, limits the global time-step $\Delta t$. 
Potential solutions to improve computational efficiency consist in the use of local time stepping techniques \cite{rossinelli2015mrag} or the implicit update of the elastic body dynamics \cite{till2017elastic}.

After providing a comprehensive explanation of our algorithm, we proceed to assess its accuracy and convergence properties through rigorous validations against a variety of analytical and numerical benchmarks.

\section{Validation benchmarks}\label{sec:bmks}

The conducted benchmark tests involve flow past a two-dimensional elastic rod under gravity, a two-dimensional elastic rod flapping in the wake of a rigid cylinder, cross-flow past a three-dimensional elastic rod under gravity, and self-propelled elastic anguilliform swimming in two and three dimensions.
For each case, we conduct a convergence analysis by reporting the discrete error \(e\) (of relevant physical quantities) as a function of the spatial discretization. 
We define \(e\) as \(e = |(p - p_{ref}) / p_{ref}|\), where \(p\) is a physical quantity obtained from our method and \( p_{ref}\) is the reference solution.

In different problems, the dynamics involved can be influenced by one or more key dimensionless numbers. 
Here, we provide a compilation of these numbers along with their respective physical interpretations
\begin{equation}
        \Rey := ~~~\frac{\rho_f V L}{\mu_f} \sim\frac{\textrm{inertial
        forces}}{\textrm{viscous forces}}; ~~~
        \Ca := ~~~\frac{\rho_f V^2}{E} \sim\frac{\textrm{inertial
        forces}}{\textrm{elastic forces}}; ~~~
        \rho := ~~~\frac{\rho_e}{\rho_f} \sim\frac{\textrm{body inertia}}{\textrm{flow inertia}}
        ~~~
        \AR := ~~~\frac{L}{D} \sim\textrm{Slenderness}
\label{eq:nondim}
\end{equation}
where \Rey, \Ca, $\rho$, \AR, $V$, $L$, $\mu_f$, $\rho_f$, $\rho_e$, $E$, $D_e$  and $L_e$ correspond to the Reynolds number, Cauchy number \cite{massey1998mechanics}, density ratio, rod's aspect ratio, velocity scale, length scale (i.e. the rod length), fluid viscosity, fluid density, elastic rod density, Young's modulus of the rod and the rod's cross-sectional diameter, respectively. 
Besides the numbers above, another dimensionless quantity has been conventionally employed to characterize the dynamics of slender body systems~\cite{turek2006proposal,silva2018determination}. 
This number, known as the dimensionless bending stiffness \(K_b\), is defined as
\begin{equation}
        K_b :=~~~\frac{EI}{\rho_f V^2 L^3}~\textrm{(2D)}~~~\textrm{or}~~~\frac{EI}{\rho_f V^2 L^3 D} ~\textrm{(3D)}~~~\sim\frac{\textrm{ bending mode forces}}{\textrm{inertial forces}}
\label{eq:nondim_bending_stiffness}
\end{equation}
where $I$ corresponds to the second area moment of inertia of the rod's cross-section about the normal/binormal axis (see \cref{sec:cosserat_rod}). The slight difference in the definition of $K_b$ in 2D and 3D stems from the difference in units of $I$, which has units $\textrm{m}^3$ in 2D and $\textrm{m}^4$ in 3D. Lastly, we note that $K_b$ can be expressed as a function of the other non-dimensional parameters mentioned in \cref{eq:nondim}. For instance, in the case of an elastic rod in 3D with a cross-sectional diameter $D$, the moment of intertia scales as $I \propto D^4$, resulting in the equivalent definition $K_b := (\Ca \AR^3)^{-1}$.

\subsection{Flapping of a two-dimensional elastic rod in the wake of a rigid cylinder}\label{sec:rod_flap_in_rigid_cyl_wake}

\begin{figure}[!ht]
    \centering
    \includegraphics[width=\textwidth]{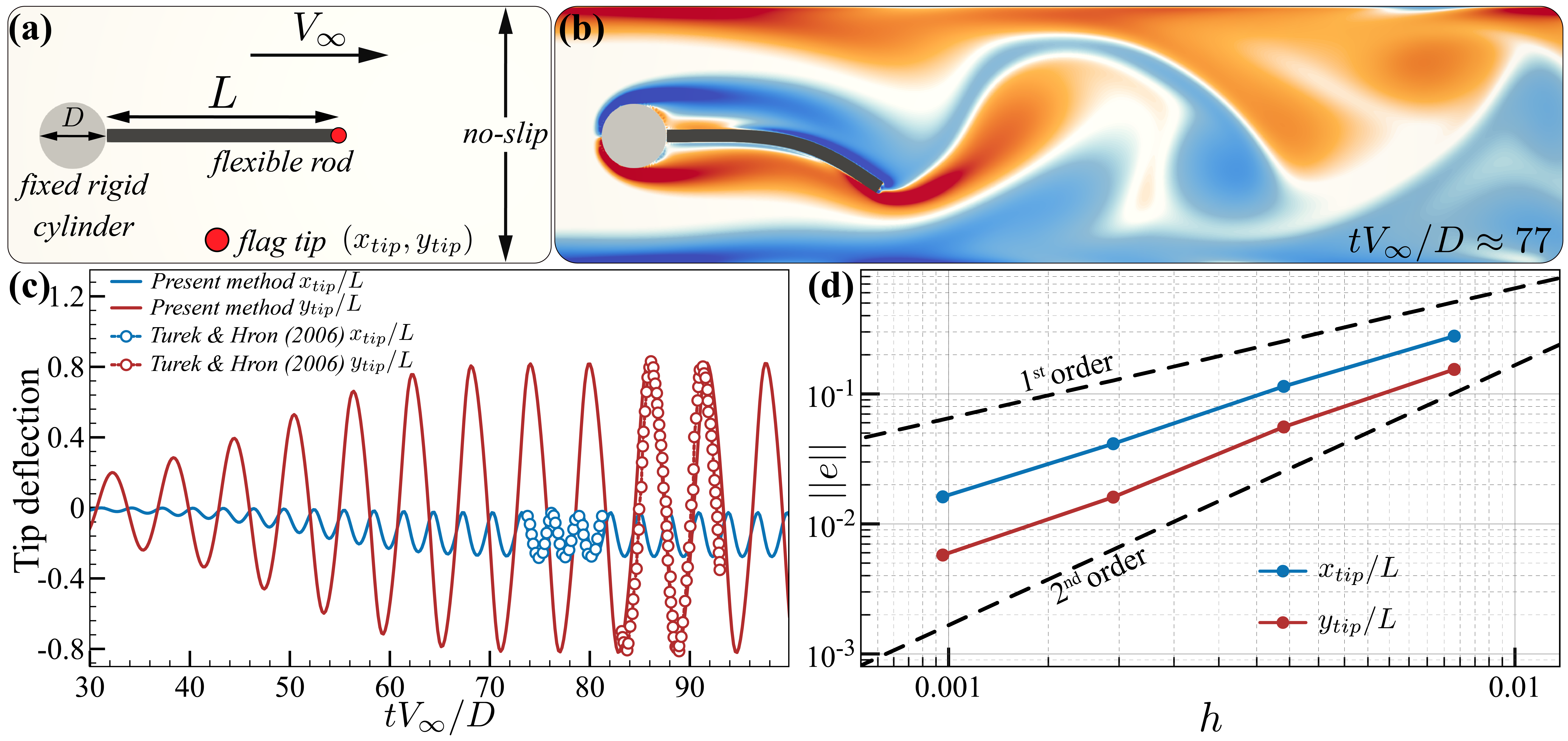}
    \caption{Flapping of a two-dimensional elastic rod in the wake of a rigid cylinder. (a) Setup. The 2D rigid cylinder of diameter $D$ is initialized inside a 2D flow domain of size $(17.5D, 6.5625D)$, with its center at $(3D, 3.28125D)$. The 2D elastic rod with length $L = 3.5D$, \(\AR = 17.5\) and mass per unit length $m_e$, is initialized with one end fixed at $(3.5D, 3.28125D)$ and centerline at $y = 3.28125D$. Both these bodies are immersed in a viscous fluid with a background free-stream velocity \(V_{\infty} \hat{\gv{i}}\). Additionally, two no-slip walls are initialized at $y = 1.28125D$ and $y = 5.28125D$, respectively. The key non-dimensional parameters in this benchmark are \(t V_{\infty} / D\), \(\Rey:= V_{\infty} D / \nu_f = 100\), \(K_b = 0.02177\) and non-dimensional mass ratio \(M_e = m_e / \rho_f L = 0.5714\). Other computational parameters are \(\CFL = 0.1\), \(\alpha = -5 \times 10^4\) and \(\beta = -20\).
    (b) Snapshot of the vorticity field (orange and blue indicate positive and negative vorticity respectively) for \(t V_{\infty} D \approx 77\), depicting the rod flapping motion and vortex shedding. 
    (c) Comparison of temporal variation of the rod tip's horizontal and vertical displacements against previous studies \cite{turek2006proposal}. 
    (d) Grid convergence. Error norms for the amplitudes of the rod tip's vertical and horizontal displacements, plotted against grid spacing $h$.
    Further details can be found under the \texttt{examples/2d_examples/VortexInducedFlagFlappingCase} folder of \cite{sopht2023}.
    }
    \label{fig:2d_turek_hron}
\end{figure}

We first demonstrate the ability of our solver to capture interactions between multiple elastic/rigid bodies immersed in a viscous fluid.
Accordingly, we reproduce the case of a two-dimensional elastic rod flapping in the wake of a rigid cylinder, characterized in detail by \citet{turek2006proposal}. 
\Cref{fig:2d_turek_hron}a presents the initial physical setup---a fixed rigid cylinder of diameter $D$ is immersed in a channel of width $4D$, with a constant, unbounded, background free stream of velocity \(V_{\infty} \hat{\gv{i}}\). 
An elastic rod of length $L$ is initialized in the same fluid, with its upstream end clamped to the most downstream point of the cylinder.
Additional geometric and parametric details can be found in \cref{fig:2d_turek_hron} caption.
We simulate this cylinder–rod system long enough after shedding vortices to eventually reach a dynamic, periodic state. 
This state is visualized via the vorticity contours at a particular time instance in \cref{fig:2d_turek_hron}b, where the rod is observed to flap and shed vortices downstream in the channel.
We then quantitatively validate our solver by tracking the temporal variation of the rod tip’s horizontal and vertical displacements and compare it against \citet{turek2006proposal}. 
As seen from \cref{fig:2d_turek_hron}c, our results show close agreement with the benchmark \cite{turek2006proposal}.

We then present the grid convergence for this case, by tracking the amplitude of the rod tip's horizontal and vertical oscillations (\cref{fig:2d_turek_hron}c), and computing the error norms with respect to the best resolved case.
With $\CFL = 0.1$, we vary the spatial resolution between $128 \times 48$ and $1024 \times 384$ (with $2048 \times 768$ as the best resolved case). 
As seen from \cref{fig:2d_turek_hron}d, we observe grid convergence between first and second order (least squares fit of 1.33 and 1.24 for vertical and horizontal displacements, respectively), consistent with our algorithm.
Thus, the results of this section indicate the ability of our approach to successfully capture, in two dimensions, fluid–elastic solid and fluid–rigid solid interactions that are themselves coupled.
For an additional, simpler validation of a single, two-dimensional elastic structure immersed in fluid, the reader is referred to the \cref{app:bmk_flow_past_2d_rod}.

\subsection{Cross-flow past a three-dimensional elastic rod under gravity}\label{sec:bmk_flow_past_3d_rod}

\begin{figure}[!ht]
    \centering
    \includegraphics[width=\textwidth]{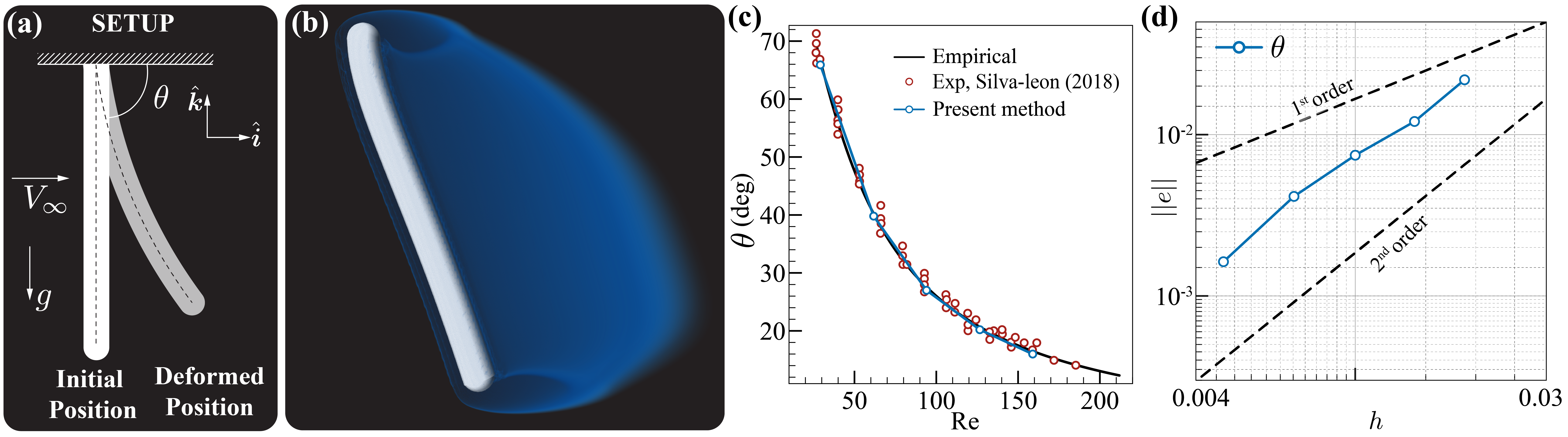}
    \caption{
    Cross-flow past a three-dimensional elastic rod under gravity.
    (a) Setup. A 3D elastic rod with length $L$,  $AR=11$, $\rho=826.45$ is initialized inside a 3D flow domain of size $(1.8L, 0.45L, 1.8L)$, with one end fixed at $(0.36L, 0.225L, 1.35L)$ and center-line along $x=0.36L$.
    The rod is immersed in a viscous flow with a free-stream velocity $V_{\infty} \hat{\gv{i}}$ and a gravitational field $-g \hat{\gv{k}}$.
    The rod deflection angle $\theta$ is measured at equilibrium, where a balance between elastic, flow and gravitational forces is present.
    The key non-dimensional parameters in this benchmark are $K_b$, Froude number $Fr:=gD/V_{\infty}^2$, stretch to bending ratio $K_s/K_b := EAL^2/(EI) = 62500$ and $Re := V_{\infty} D / \nu_f$. 
    Other computational parameters are  CFL=0.1, $\alpha=-2 \times 10^4$, and $\beta=-10^{2}$. 
    (b) Snapshot of the steady state deformation of the elastic rod with volume rendering of vorticity magnitude (lighter and darker blue indicate lower and higher vorticity magnitude, respectively).
    The physical parameters for this simulation are set to $Re=29.139$, $K_b=0.0309$, and $Fr=3.24\times 10^{-3}$. 
    The spatial resolution of the domain is set to $256 \times 256 \times 64$.
    (c) Comparison of rod deflection $\theta$ vs $Re$ against empirical (black line) \cite{Silva:2018} and experimental (red circles) \cite{Silva:2018} results.
    In order to be consistent with experiments, we vary $K_b$ as $K_b=534.47/Re^2$ and $Fr$ as $Fr=2.75/Re^2$.
    (d) Grid convergence. Error norm of rod deflection $\theta$ for $Re=29.139$ plotted against grid spacing $h$.
    Further details can be found under the \texttt{examples/3d_examples/FlowPastRodCase} folder of \cite{sopht2023}.
    }
    \label{fig:3d_validation_flow_past_rod}
\end{figure}

Post validation in two dimensions.
Here, we consider the case of cross-flow past a three-dimensional elastic rod suspended under gravity, characterized experimentally by \citet{Silva:2018}.
\Cref{fig:3d_validation_flow_past_rod}a presents the initial setup---a three-dimensional elastic rod is immersed in a constant, unbounded, background free stream of velocity \(V_{\infty} \hat{\gv{i}}\) and is subject to the gravity field \(-g \hat{\gv{k}}\). 
The rod is clamped at its top end, allowing it to deform in response to the surrounding cross-flow. 
Computational setup details can be found in \cref{fig:3d_validation_flow_past_rod}.
We simulate the system long enough until the rod deforms and attains a steady deformed equilibrium state, where the drag forces due to cross-flow are balanced by gravity and by the restoring internal stresses. 
This equilibrium state of the rod is visualized via  volume rendered vorticity magnitude, for a particular cross-flow speed, in \cref{fig:3d_validation_flow_past_rod}b.
Next, for quantitative validation, we characterize the deformed state via $\theta$, the angle between the axis of the rod and the direction of the free stream flow. 
We then measure the deformation angle $\theta$ for different flow speeds (which correspond to different Reynolds numbers $\Rey$) and compare it against the experiments of \citet{Silva:2018}.
As seen from \cref{fig:3d_validation_flow_past_rod}c, our results agree with the benchmark \cite{Silva:2018}.

Next, we present the grid convergence for this case, by tracking the deformation angle $\theta$ for a particular free stream speed ($\Rey=29.139$), and computing the error norms against the best resolved case.
With $\CFL = 0.1$, we vary the spatial resolution between $96 \times 96 \times 24$ and $384 \times 384 \times 96$ (with $512 \times 128 \times 128$ as the best resolved case). 
As seen from \cref{fig:3d_validation_flow_past_rod}d, our method presents convergence between first and second order (least square fit of 1.81), consistent with the proposed algorithm.

\subsection{Self-propelled soft anguilliform swimmers}\label{sec:bmk_fish}

\begin{figure}
    \centering
\includegraphics[width=\textwidth]{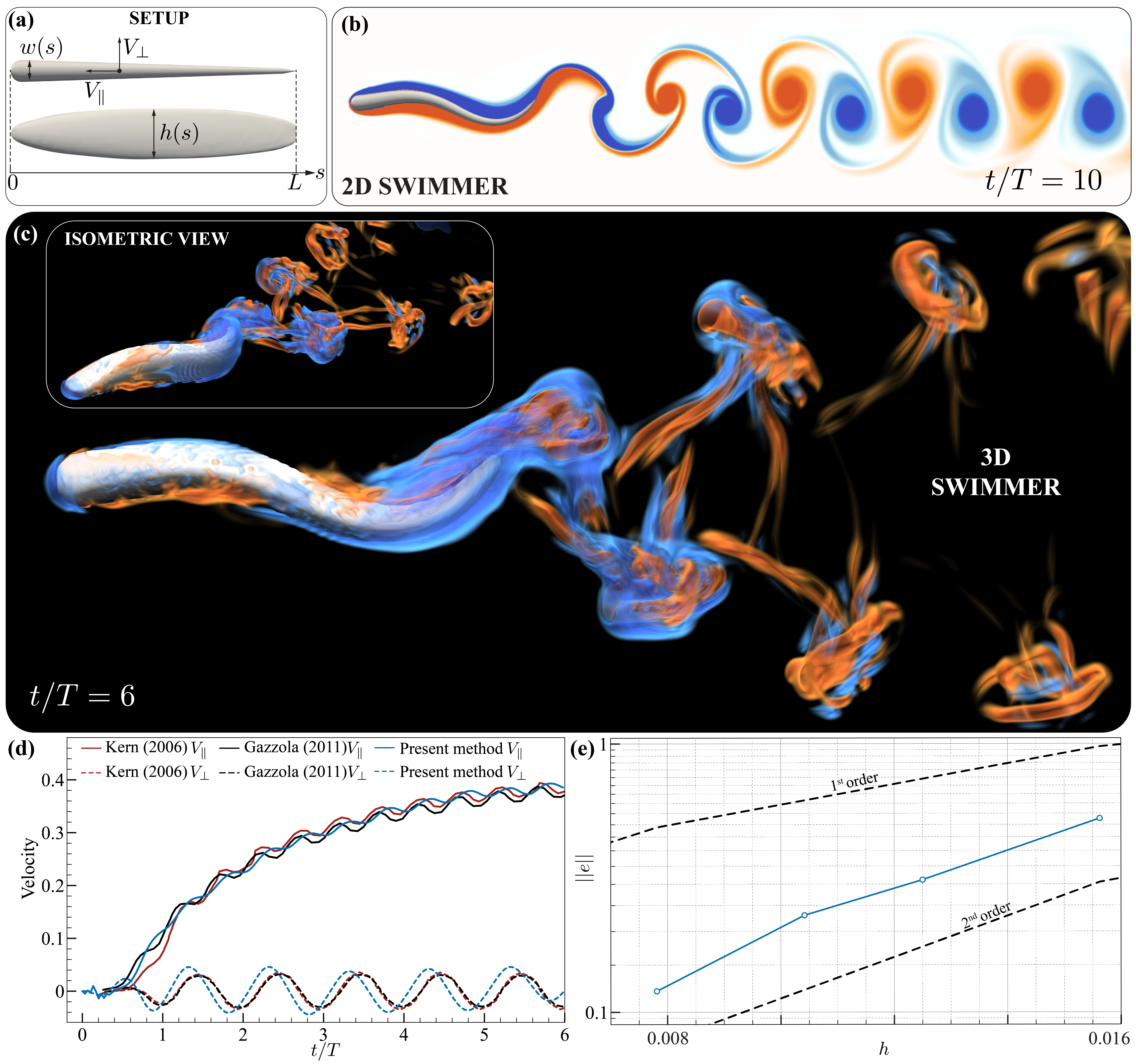}
    \vspace{-0.75cm}
    \caption{
    Self-propelled elastic anguilliform swimmers in two and three dimensions.
    (a) Setup.
    An elastic swimmer with length $L$, arc-length $s \in [0,L]$, half width $w(s)$ and half height $h(s)$ is immersed in an unbounded flow domain and is actuated using forces and torques that generate the periodic mid-line kinematics (with time-period $T$) described in \cref{eq:carling_kinetics}.
    Here, we consider two- and three-dimensional variants of the swimmer, where $w(s)$ defines the geometry in 2D, while $w(s)$ and $h(s)$ define the geometry of the body in 3D (see \cref{sec:fish_geometry}).
    In order to capture the effect of varying cross-sections on the swimmer dynamics, we modify $I_{i}$ (the second moment of the area around the $i$-director) accordingly along the rod. 
    In 2D, the elastic swimmer is initialized with its head ($s=0$) at $\left(5.45L, 0.68L\right)$ in a flow domain of size $\left(7L, 1.38L\right)$.
    In 3D, the body head is initialized at $\left(2.9L, L, 0.25L\right)$ in a flow domain of size $\left(4L, 2.0L, 0.5L\right)$.
    The elastic swimmer's motion is characterized via the velocity components of the center of mass, that are parallel ($V_{\parallel}$) and perpendicular ($V_{\perp}$) to the swimming direction.
    These velocity components ($V_{\parallel}$ and $V_{\perp}$) are normalized by the velocity scale $L / T$.
    Key non-dimensional parameters in this benchmark are $t/ T$, $\rho=1$, actuation Reynolds number $Re := L^2/(T \nu_f)=7142$, $K_b$ (set as 0.7044 in 2D and 0.1724 in 3D).
    Other computational parameters are $\CFL=0.1$, \(\alpha\) (set as $-8 \times 10^4$ in 2D and $-3.2 \times 10^4$ in 3D) and \(\beta\) (set as $-10$ in 2D and $-32$ in 3D). 
    (b) Snapshot of the vorticity field (orange and blue indicate positive and negative vorticity, respectively) for $t/T=10$, depicting the wake generated by the 2D soft swimmer. 
    The spatial resolution of the 2D domain for the visualized case is set as $3072 \times 614$.
    (c) Snapshot of the wake generated by the 3D soft swimmer with the volume rendered Q-criterion (orange and blue indicate vorticity and strain dominated areas, respectively) for $t/T=6$. 
    Inset on the top left corner shows the isometric view of the snapshot. 
    The spatial resolution of the 3D domain for the visualized case is set as $768 \times 384 \times 96$.
    (d) Validation. Comparison of temporal variation of the elastic swimmer's velocity components ($V_{\parallel}$ and $V_{\perp}$) against previous studies \cite{Kern:2006,gazzola2011simulations} for the 3D swimmer. Validation for the 2D swimmer can be found in \cref{fig:2d_validation_fish_swimming}a.
    (e) Grid convergence. Error norm of $V_{\parallel}$ plotted against grid spacing $h$ for the 3D swimmer. Convergence study of the 2D swimmer can be found in \cref{fig:2d_validation_fish_swimming}b.
    Further details can be found under the \texttt{examples/2d_examples/FishSwimmingCase} folder for \blue{the} 2D swimmer and the \texttt{examples/3d_examples/FishSwimmingCase} folder for the 3D swimmer of \cite{sopht2023}.
    }  \label{fig:3d_validation_fish_swimming}
\end{figure}

Following the successful validation of our technique against a three-dimensional static benchmark, we now demonstrate its ability to capture untethered, three-dimensional coupled dynamics, via the case of a self-propelled anguilliform swimmer immersed in a viscous fluid.
We point out that this benchmark is the most demanding of those illustrated so far. 
This is because the validation depends on precisely capturing the dynamics of active, unconstrained slender bodies with varying cross-sections, while simultaneously resolving the interplay between body inertia, elasticity, viscous drag and flow inertia.

In \Cref{fig:3d_validation_fish_swimming}a we presents the physical setup of the anguilliform swimmer, as described by \citet{Kern:2006}. The swimmer has length $L$, a cross-section that varies along the centerline coordinate $s \in [0,L]$, and is immersed in an unbounded fluid of the same density. 
The body geometry of the swimmer is described by the half-width $w(s)$ in two dimensions, and by the half-width $w(s)$ and half-height $h(s)$ in three dimensions (see \cref{sec:fish_geometry}).
In \citet{Kern:2006}, the swimmer's deformation (gait) is kinematically imposed, i.e. the influence of the fluid on the fish only affects rigid body translational and rotational velocity components. Here instead, the body of the fish is an elastic Cosserat rod, fully coupled to the flow and actuated using a traveling wave of torques and forces, modeling muscular actuation \cite{Gazzola:2018}. 
Here, torques and forces are back-calculated so as to reproduce the body lateral undulations of \citet{Kern:2006}, in turn based on \citet{Carling:1998}. For this gait, the lateral displacement of the mid-line $y(s,t)$, is defined as
\begin{equation}
y(s,t)=0.125L\frac{\displaystyle0.03125+\frac{s}{L}}{1.03125}\sin\left[2\pi\left(\frac{s}{L}-\frac{t}{T}\right)\right]
\label{eq:carling_kinetics}
\end{equation}
\noindent
where $t$ is the time and $T$ is the period. 
Additional computational details can be found in the caption of \cref{fig:3d_validation_fish_swimming}, while the procedure to determine the 
actuation parameters are reported in \cref{sec:fish_actuation}.

In all simulations, the elastic swimmer starts from a straight, rest configuration and actuation forces and torques are gradually increased from zero to their designated values during the first cycle.
We simulate the swimmers long enough until they reach a steady swimming average terminal speed.
For a two-dimensional swimmer, the vorticity contours at a particular time instance are reported in \cref{fig:3d_validation_fish_swimming}b.
In three dimensions, vortex rings are observed in the swimmer's wake, visualized via the volume rendered Q-criterion at a particular time instance (\cref{fig:3d_validation_fish_swimming}c).
The wakes of both 2D and 3D swimmers are found to be qualitatively consistent with benchmark literature \cite{gazzola2011simulations,Kern:2006}. Next, we proceed with quantitatively validating our simulations by characterizing
the swimmers' motion via the velocity components of the center of mass (normalized by the velocity scale $L / T$) that are parallel ($V_{\parallel}$) and perpendicular ($V_{\perp}$) to the swimming direction.
We track the temporal variation of the velocity components \(V_{\parallel} \text{ and } V_{\perp}\) and compare them against previous studies \cite{gazzola2011simulations,Kern:2006}.
As seen from \cref{fig:3d_validation_fish_swimming}d for the three-dimensional case, our results show close agreement with the benchmarks.
The two-dimensional quantitative comparison is instead reported in the \cref{sec:2d_fish_validation_section} for brevity, albeit we note similar good agreement as for the 3D case.
Finally, we present the grid convergence by tracking the terminal velocity components \(V_{\parallel} \text{ and } V_{\perp}\) and computing the error norms against the best resolved case. 
For three-dimensional swimming, with $\CFL = 0.1$, we vary the spatial resolution between $256 \times 128 \times 32$ and $512 \times 256 \times 64$ (with $768 \times 384 \times 96$ as the best resolved case).
As seen from \cref{fig:3d_validation_fish_swimming}e, our method presents convergence between first and second order (least squares fit 
of 1.16 for $V_{\perp}$ and
2.1 for $V_{\parallel}$), again consistent with the spatial discretization employed in our algorithm.

Overall the results of this Section validate our algorithm against a range of benchmarks, showing the accuracy and robustness of our numerical scheme and its implementation. 
These results are complemented by a convergence analysis which is found to be consistent with the employed discrete operators and across physical scenarios. 

\section{Multi-physics illustrations}\label{sec:ills}

Here, we further highlight the versatility of our solver by demonstrating a range 
of multi-physics applications. 
These include flow past a three-dimensional soft rod in an inertia dominated flow environment, viscous streaming induced by magnetic soft assemblies of filaments, and the locomotion of a self-propelling soft octopus.
Before proceeding with the illustrations, we recall the key dimensionless parameters \cref{eq:nondim,eq:nondim_bending_stiffness} that govern the dynamics of these systems, along with their physical interpretations---\Rey~(Reynolds number) defined as the ratio of inertial to viscous forces, \Ca~(Cauchy number) as the ratio of inertial to elastic forces, $\rho$~(density ratio) as the ratio of body inertia to flow inertia, $AR$~(aspect ratio) as the ratio of body length to diameter, and $K_b$~(bending stiffness) as the ratio of bending forces to inertial forces.

\subsection{Three-dimensional flapping of an elastic rod}\label{sec:rotating-rod}

For this demonstration we consider the three-dimensional flapping of an elastic rod due to an inertia dominated flow characterized by $\Rey \sim \order{10^3}$. 
Such inertia dominated flow regimes typically present fine spatio-temporal flow features, which necessitate high-resolution simulations to accurately resolve the system dynamics. 
To enable high-resolution simulations, we implement a parallel MPI (Message Passing Interface) based version~\cite{sopht-mpi:2023} of our algorithm that maps onto distributed computing architectures (for implementation details see \cref{app:mpi-extension}). 
Through this case we then demonstrate how our parallel implementation scales effectively across state-of-the-art distributed computing systems.

\begin{figure}[!ht]
    \centering
    \includegraphics[width=\textwidth]{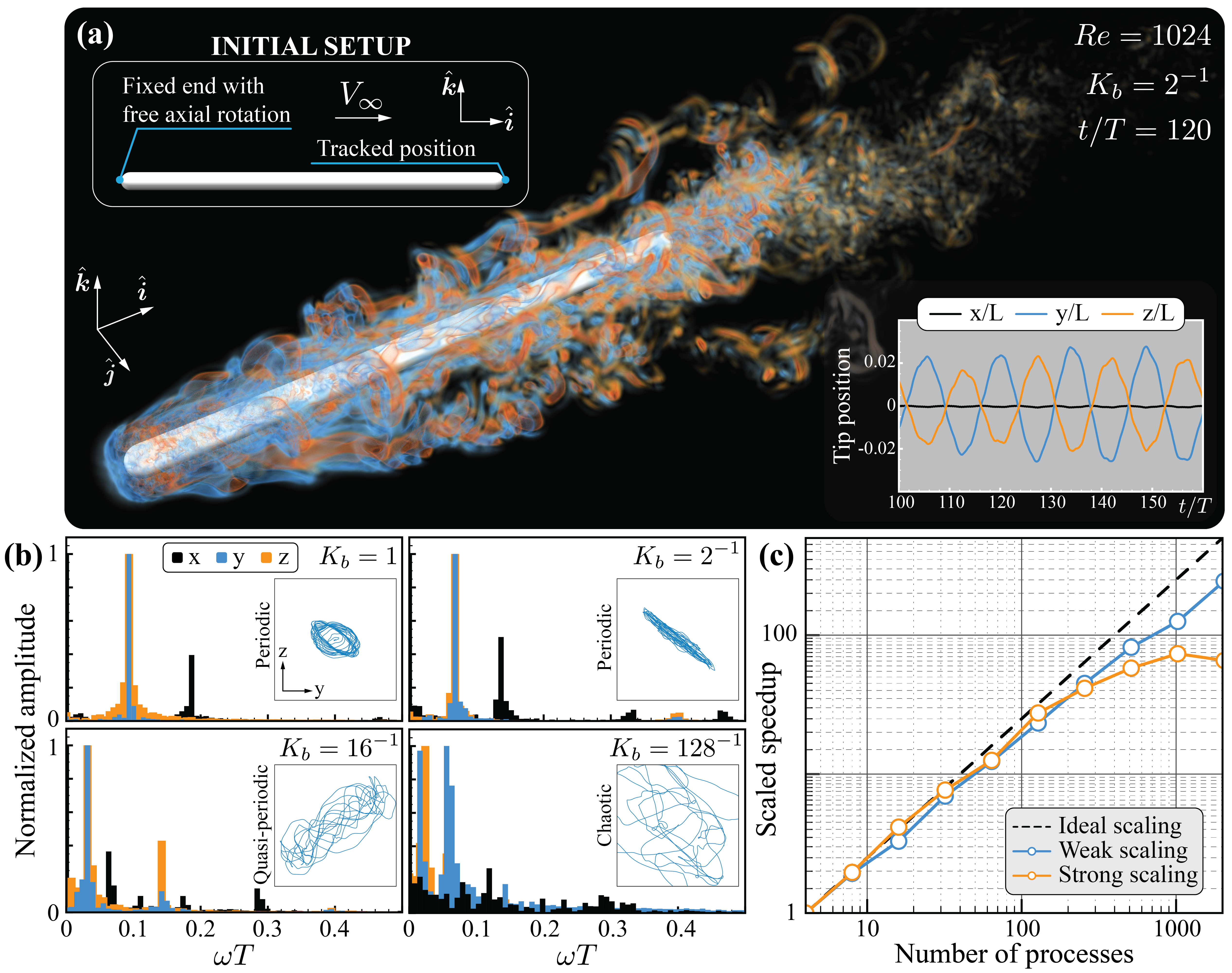}
    \vspace{-0.75cm}
    \caption{
        Flow past a three-dimensional elastic rod in an inertial flow regime.
        (a) Setup (top-left inset). A soft Cosserat rod of length $L$, $AR = 20$, \(\rho = 1000\) is initialized inside a 3D flow domain of size \((5L, 1.25L, 1.25L)\), with one freely rotating end (along its axis) clamped at \((0.625L, 2.51L, 2.51L)\) and with centerline along \(x = 0.625L\).
        The rod is immersed in a viscous flow with a free-stream velocity \(V_{\infty} \hat{\gv{i}}\).
        The key non-dimensional parameters in this illustration are \(t/ T = t V_{\infty} / L\), \(\Rey:= V_{\infty} L / \nu_f\) and \(K_b\). 
        Other computational parameters are \(\CFL = 0.1\), \(\alpha = -2 \times 10^4\) and \(\beta = 10\).
        Snapshot of the flow generated by the flapping rod visualized via the volume rendered Q-criterion (orange and blue indicate vorticity and strain dominated areas, respectively) for $t/T = 120$.
        The trajectory of the free end tip position is tracked over time as shown in the bottom right inset.
        The spatial resolution of the 3D domain is set to $1024 \times 256 \times 256$.
        (b) Fourier analysis of the rod tip's position history for varying \(K_b\), illustrating the emergence of additional frequency modes as \(K_b\) decreases. 
        The appearance of these additional modes corresponds to the system transitioning from periodic to quasi-periodic to finally chaotic behaviour, as qualitatively seen in the traverse tip deflection visualized in the insets (the axes are set to range from -0.05L to 0.05L). 
        (c) Plot of the scaled speedup against the number of employed processes, illustrating the weak and strong scalability of our algorithm when mapped onto distributed parallel architectures.
        For weak scaling, we set $32^3$ as the spatial resolution per process, and record the timings as a function of the number of processes. 
        For strong scaling we set the spatial resolution to $384^3$, and record the timings with varying number of processes.
        The timing related simulations are conducted on Stampede2 KNL nodes \cite{Stanzione:2017}, and the reported speedups are obtained by scaling with the $\mathcal{O}(n \log n)$ complexity of Fourier-based Poisson solve operation \cite{sbalzarini2006ppm}.
        Further details can be found under the \texttt{examples/3d_examples/FlowPastFreelyRotatingRod} folder of \cite{sopht-mpi:2023}.
    }
    \label{fig:rotating_rod}
\end{figure}

We begin, as shown in \cref{fig:rotating_rod}a (upper left inset), by initializing a three-dimensional elastic rod immersed in a constant, unbounded, background free stream of velocity \(V_{\infty} \hat{\gv{i}}\). 
The rod is clamped at its upstream end and is free to rotate about its axis, thus allowing flapping and rotation in response to the surrounding flow.
Additional geometric and parametric details can be found in \cref{fig:rotating_rod}.
We simulate this system long enough until the rod sheds vortices, and eventually reaches either a periodic, quasi-periodic or chaotic state, depending on the rod elasticity $K_b$.
This state is visualized via the volume rendered Q-criterion at a particular time instance and $K_b$, in \cref{fig:rotating_rod}a.
To characterize the dynamical behavior of the system, we simulate it for varying rod stiffness \(K_b\) and track the coordinate of the rod's free tip (marked in \cref{fig:rotating_rod}a, upper left inset) over time (lower right inset of \cref{fig:rotating_rod}a).
In \cref{fig:rotating_rod}b, the Fourier analysis of the rod tip's position history is presented for different values of $K_b$, with the rod's lateral displacement trajectories shown in the insets. 
When the rod is stiff (represented by higher values of $K_b$, shown in the top two plots of \cref{fig:rotating_rod}b), its dynamics are periodic, characterized by a single peak in the Fourier analysis and a periodic limit cycle in the tip trajectory insets. 
As the rod becomes softer (i.e. $K_b$ decreases), additional frequency modes emerge in the Fourier spectrum, represented by additional smaller peaks in the Fourier analysis (bottom left plot), and the system transitions from periodic to quasi-periodic dynamics. 
This is reflected as a quasi-cycle (an approximate cycle that does not repeat exactly) in the trajectory inset. 
Further reduction in the rod's stiffness $K_b$ results in a transition to chaotic dynamics, demonstrated by a continuous frequency spectrum in the Fourier space (bottom right plot), and chaotic non-repeating curves in the tip trajectory inset.
Previous reports have indicated similar dynamical transitions in flapping two-dimensional filaments \cite{alben2008flapping,chen2014bifurcation}, but no study has investigated three-dimensional counterparts. 
Although a complete analysis of this system lies beyond the scope of the current paper, this illustration showcases how our algorithm, combined with high-performance computing strategies, can be employed to explore uncharted systems requiring high fidelity simulations, with potential applications in energy harvesting, drag reduction, and related fields.

We conclude this investigation by highlighting how the parallel implementation of our algorithm scales effectively across multiple distributed computing systems.
In \cref{fig:rotating_rod}c, we present the speedup values for both weak (scaling domain size with the number of processes) and strong (keeping the problem size fixed while varying the number of processes) scaling, as a function of the number of employed processes (for details see \cref{fig:rotating_rod} caption). 
The results depicted in \cref{fig:rotating_rod}c demonstrate that the weak and strong scaling trends align closely with the ideal theoretical trend up to approximately one thousand employed processes, underscoring the efficiency of our algorithm, on top of its accuracy, versatility and relative simplicity.

\subsection{Streaming flow generation via magnetized filament assemblies}\label{sec:cilia}

\begin{figure}[!ht]
    \centering
    \includegraphics[width=0.925\textwidth]{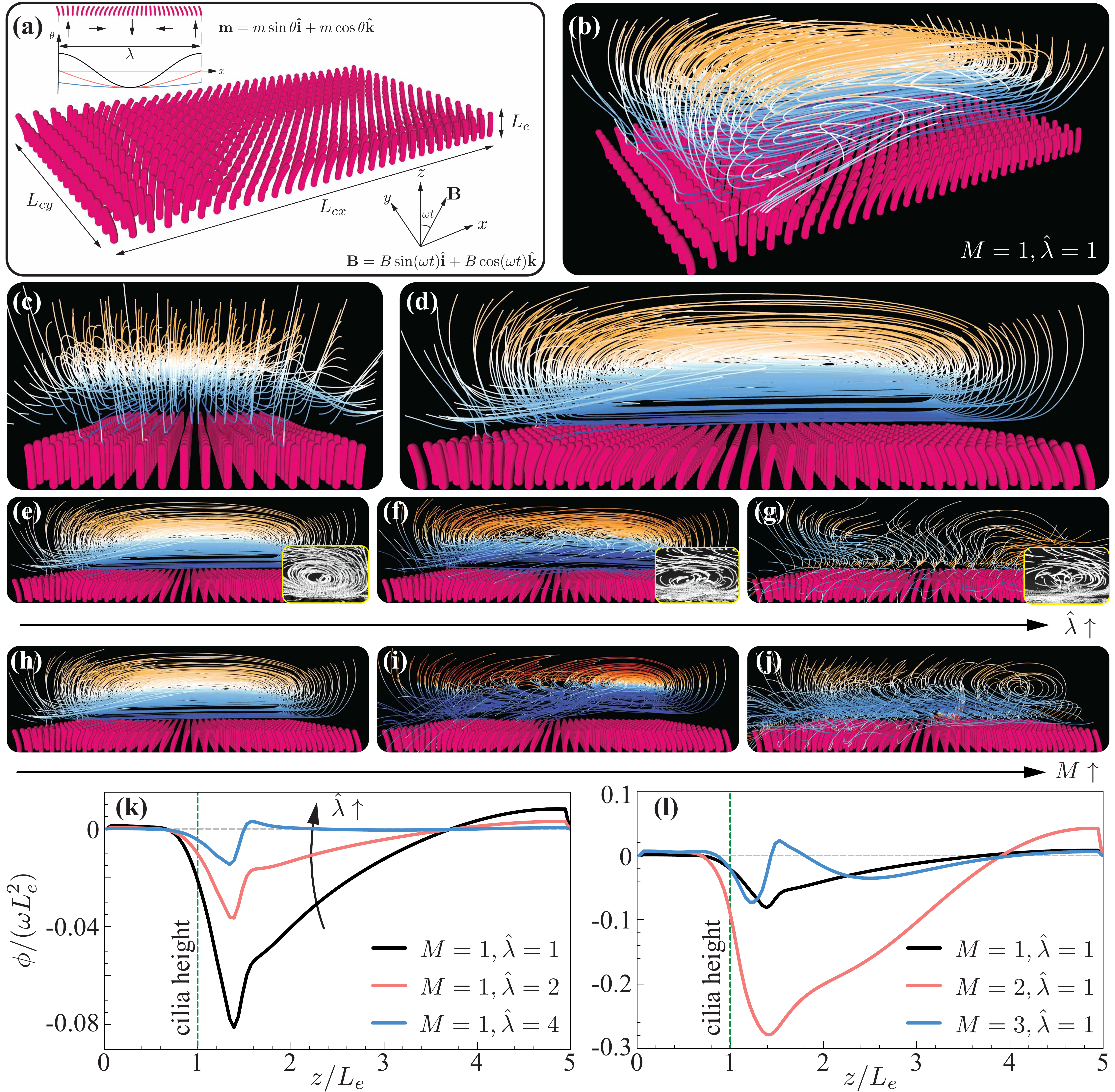}
    \caption{
    Streaming flow generation via magnetized filament assemblies. 
    (a) Setup. A three-dimensional magnetic cilia carpet, composed of individual elastic rods of length $L_e$, and arranged in a rectangular uniform grid of 32 by 16 rods is initialized in a three-dimensional flow domain. 
    The carpet has a span of $L_{cx}$ and $L_{cy}$ along the $x$ and $y$ directions, respectively, with a rod-to-rod spacing of $L_e/2$ along $x$ and $y$ axes. Each filament, with $AR=5$ and $\rho = 2390$, is initialized with the lower end fixed, and the centerline pointing along the $z$ direction. 
    The fluid domain is of size ($L_{cx} + 2L_e$, $L_{cy} + 2L_e$, 5$L_e$) with the spatial resolution set as $384\times208\times110$ throughout.
    Similar to the experimental setting of \citet{gu2020magnetic}, the cilia are strongly magnetized with magnetic dipole moment density $\gv{m}$, where $\gv{m}=m\sin\theta \gv{i} + m\cos\theta \gv{k}$. Here the magnetization angle $\theta$ varies spatially along the $x$ direction following an antiplectic pattern with non-dimensionalized wavelength $\hat{\lambda} = \lambda / L_{cx}$. 
    Following setup, the cilia carpet is subjected to a weak, oscillatory magnetic field of strength $\gv{B}$ that rotates sinusoidally in the $x-z$ plane given by $\gv{B} = B\sin(\omega t) \gv{i} + B\cos(\omega t) \gv{k}$. 
    The key non-dimensional parameters in this benchmark are the Womerseley number $\Mn:= L_e \sqrt{\omega / \nu_f}$, magnetic Bond number $Bo:= (mBAL^2)/(EI) = 3.0$ (ratio of magnetic to elastic forces), and the non-dimensional oscillation frequency of the magnetic field $\omega / \omega_0 = 0.16926$, where $\omega_0 = 3.5160 / L_e^2 \sqrt{EI/(\rho_e A)}$ is the natural frequency of the filament's first bending mode. 
    Other computational parameters are CFL=0.1, $\alpha=-2 \times 10^4$, and $\beta=-10$.
    The rotating magnetic field creates a metachronal wave that propagates in the $x$-direction along the cilia carpet, ultimately generating streaming flow on top of the carpet. 
    Streaming flow topology visualized via the Eulerian time-averaged streamlines for $M = 1$ and $\hat{\lambda} = 1$ (orange and blue represent positive and negative $x$-velocities, respectively) in (b) an isometric view (c) $y-z$ plane and (d) $x-z$ plane, respectively. 
    (e-g) Flow topology variation with $\Mn = 1$ and increasing wavelength $\hat{\lambda}$, with (e) $\hat{\lambda} = 1$ (f) $\hat{\lambda} = 2$ and (g) $\hat{\lambda} = 4$.
    Streaming eddies visualized via particle trajectories from equivalent experiments of \citet{gu2020magnetic} (with one-to-one $\hat{\lambda}$ values) are included as yellow inset boxes for reference. 
    (h-j) Flow topology variation with $\hat{\lambda} = 1$ and increasing $\Mn$, with (h) $\Mn=1$ (i) $\Mn=2$ and (j) $\Mn=3$. 
    (k-l) Non-dimensional fluid transport flux $\phi / \omega L_e^2$ (\cref{eqn:fluid_transport_flux}) along $x$ axis (measured at the carpet midplane $x=L_e/2$) plotted against $z$ coordinate, for varying (k) $\hat{\lambda}$ and (l) $\Mn$.
    Further details can be found under the \texttt{examples/3d_examples/MagneticCiliaCarpetCase} folder of \cite{sopht2023}.}
    \label{fig:cilia_carpet}
\end{figure}

Here we demonstrate our solver's ability to capture second-order flow physics and rectification, via the example of viscous streaming generated by magnetized filament assemblies.
Viscous streaming, an inertial phenomenon, refers to the steady, time-averaged flow that emerges when a immersed body undergoes small-amplitude oscillations within a viscous fluid. 
Due to its ability to reconfigure flow topology over short time and length scales, viscous streaming has found applications in multiple aspects of microfluidics, from particle manipulation and drug delivery to chemical mixing \cite{lutz2003microfluidics,marmottant2003controlled,wang2011size,chong2013inertial,thameem2017fast,parthasarathy2019streaming,bhosale2022multicurvature}.
Despite comprehensive understanding of rigid body streaming \cite{bhosale_parthasarathy_gazzola_2020,chan2022three,bhosale2022multicurvature}, little is known when body elasticity is involved \cite{cui2023three,bhosale2022streaming,dou2022vitro}, and even less so when multiple soft streaming bodies are involved.
Yet, the rectification of flow through systems comprised of several soft components is pertinent to various biological contexts, including flows produced by ciliated organisms like bacteria and larvae \cite{gilpin2020multiscale}.
Cilia, slender hair-like structures located on the outer surface of the organism,
drive these flows typically by deforming with a spatial phase difference, termed metachrony, 
 so as to form 
wave-like motions \cite{brennen1977fluid} 
Metachronal waves have then been hypothesized to generate, in some cases, 
streaming flows 
for feeding or locomotion \cite{gilpin2020multiscale}.
Thus, motivated by the potential utility of filament arrays for streaming flow manipulation, 
we explore here the possible use of synthetic, 
magnetically driven cilia carpets, as detailed in the study by \citet{gu2020magnetic}. 

Before proceeding, we first describe the computational details involved in the modeling of magnetic filaments. 
We assume that each filament carries within its local frame a homogeneous and permanent magnetization, characterized by the magnetic dipole moment density $\gv{m}$.
An external, driving magnetic field, characterized by magnetic flux density $\gv{B}$, is assumed to be unidirectional and spatially homogeneous (i.e. gradients are identically zero). 
Then, at any time instant, the magnetic field generates an external torque $\gv{c}$ along the rod, which is described by 
\begin{equation}
\label{eqn:magnetic_torque}
    \gv{c} = \bv{Q}^T \gv{m}~\delta{V} \times \gv{B}
\end{equation}
where $\delta{V}$ is the local differential volume and $\bv{Q}$ is the local frame of reference. For details concerning the numerical implementation of the magnetic torques, the reader is referred to \cite{magnetopyelastica2023}.

Next, we adopt the physical setup shown in \cref{fig:cilia_carpet}a, where a uniform, rectangular arrangement of cilia carpet with $32 \times 16$ rods of length $L_e$ is immersed in an unbounded viscous fluid.
Similar to \citet{gu2020magnetic}, the cilia carpet is strongly magnetized with phases varying in a harmonic fashion along the $x$ direction, with a non-dimensional wavelength $\hat{\lambda} = \lambda / L_{cx}$, where $L_{cx}$ is the carpet span along the $x$ axis. 
Further details concerning the simulation setup, are reported in the caption of \cref{fig:cilia_carpet}.
The carpet is subjected to a weak, sinusoidally rotating external magnetic field of angular frequency $\omega$ in the $x$-$z$ plane, which creates a metachronal wave that propagates along the $x$ direction (see \cref{app:videos}). 
As can be seen in \cref{fig:cilia_carpet},
this metachronal wave is indeed found to generate a steady streaming response in the surrounding fluid.
Following classical streaming literature \cite{bertelsen1973nonlinear}, we characterize this response via the Womersley number $\Mn$ defined as $\Mn:= L_e \sqrt{\omega / \nu_f}$. 
The streaming flow response for the base case of $\Mn = 1$ and $\hat{\lambda} = 1$ is visualized via time-averaged Eulerian streamlines in \cref{fig:cilia_carpet}b-d. This illustrates that a steady, centered vortex appears above the cilia carpet, consistent with the experiments of \citet{gu2020magnetic}.

Next, we examine the effect of varying Womersley number ($\Mn$) and metachronal wavelength $\hat{\lambda}$ on the streaming flow topology. 
We show in \cref{fig:cilia_carpet}(e-g) the time-averaged Eulerian streamlines for a fixed $\Mn = 1$ and varying non-dimensional wavelength $\hat{\lambda} = 1$, $\hat{\lambda} = 2$, and $\hat{\lambda} = 4$. 
The wavelength captures the synchronization among the cilia, with $\hat{\lambda} \rightarrow \infty$ indicating synchronous motion across all cilia. 
With increasing $\hat{\lambda}$, we observe the dissolution of the well-defined, centered vortex, consistent with observations made in experiments (shown as insets) with equivalent wavelengths \cite{gu2020magnetic}.
We subsequently consider  
a fixed wavelength $\hat{\lambda} = 1$ and a varying Womersley number $\Mn = 1$, $\Mn = 2$, and $\Mn = 3$ (\cref{fig:cilia_carpet}h-j). 
This corresponds to a transition of the flow from a viscosity-dominated to an inertia-dominated regime. 
We observe that with increasing $\Mn$, the central vortex shifts in a direction opposite to the metachronal wave propagation direction, eventually breaking down into a collection of smaller eddies.

Finally, we characterize quantitatively the streaming flow strength via the non-dimensional fluid transport flux $\phi / (\omega L_e^2)$ in the $x$ direction, measured at the central $y$-$z$ plane of the fluid domain. The dimensional flux $\phi$ is defined as
\begin{equation}
\label{eqn:fluid_transport_flux}
    \phi(z) = \int_0^{L_{cy}} v_x(y, z)~dy
\end{equation}
where $L_{cy}$ is the carpet span along the $y$ axis, and $v_x(y, z)$ is the flow speed along $x$ axis at $x = L_{cx} / 2$.
We then track $\phi$ as a function of distance above the carpet ($z$) with varying wavelength $\hat{\lambda}$ (\cref{fig:cilia_carpet}k) and Womersley number $\Mn$ (\cref{fig:cilia_carpet}l).
Consistent with findings of \citet{gu2020magnetic}, the flow strength decreases with increasing $\hat{\lambda}$, implying diminishing streaming effects.
In contrast, varying $\Mn$ (keeping $\hat{\lambda}=1$ fixed) produces stronger non-linearities, and a local maximum of the transport flux near $\Mn = 2$. 
Overall, the above-demonstrated generation and tunability of streaming flows via artificial filament arrays hints at their potential in microfluidic applications, from particle manipulation and transport to chemical mixing~\cite{sahadevan2022microfluidic}, 
while underscoring the ability of our solver to capture such complex second order effects, in agreement with experiments.

\subsection{Octopus swimming}\label{sec:octopus}

\begin{figure}[!ht]
    \centering
    \includegraphics[width=\textwidth]{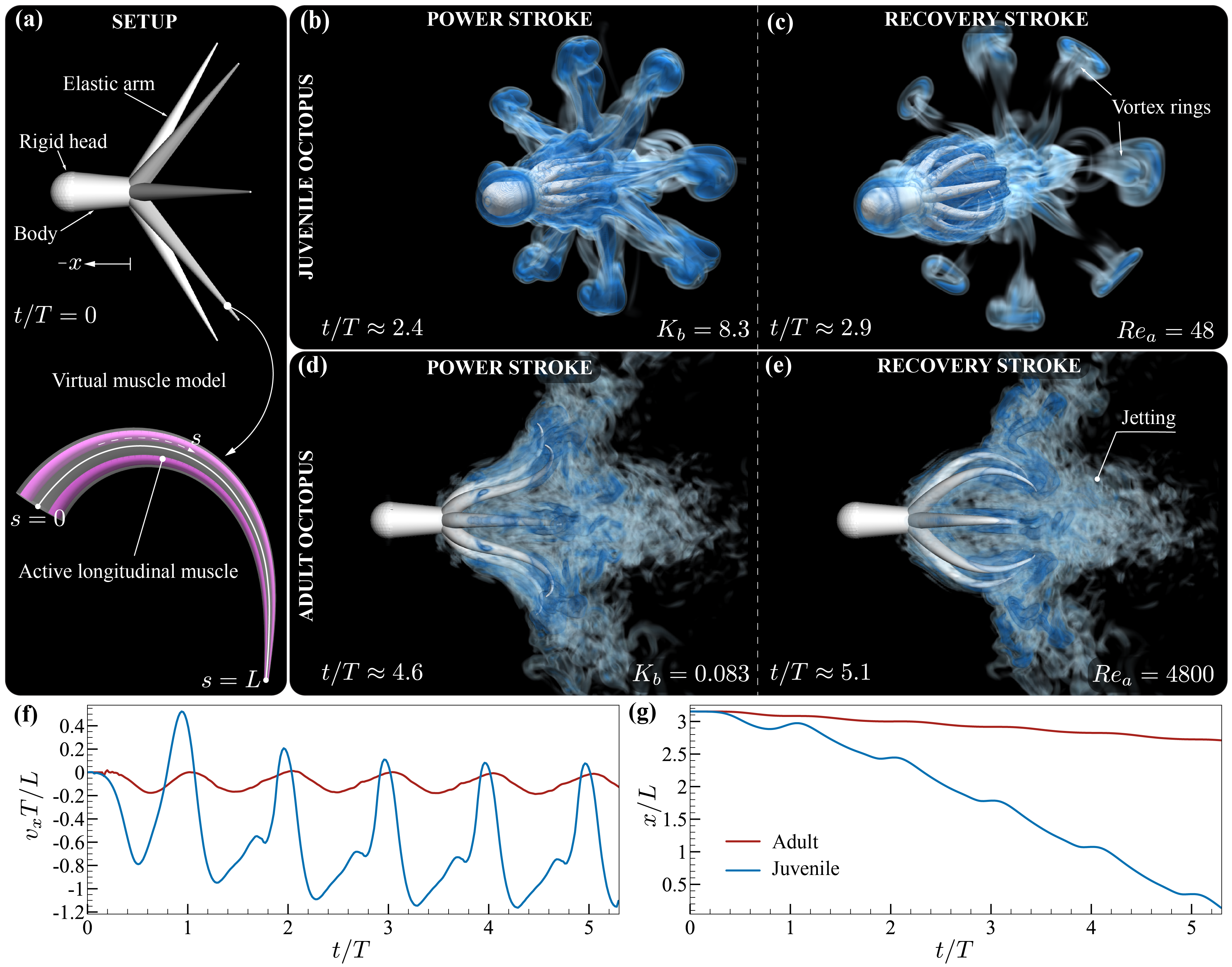}
    \vspace{-0.75cm}
    \caption{
    Octopus swimming.
    (a) Setup. An eight armed octopus with soft, tapered arms is immersed in a three-dimensional, unbounded, viscous flow. 
    Each soft arm is modeled as a Cosserat rod with length $L$, arc-length $s \in [0, L]$, radius $r(s) \in [r_{b}, r_{t}]$ (at base $r(0) = r_{b}$ and at tip $r(L)=r_{t}$), taper ratio $r_{b}/r_{t} = 12$, base $AR:= L/(2r_{b})=4.16$, and is actuated via muscular forces and torques.
    These forces and torques are computed using the virtual muscle model, where two virtual longitudinal muscles (depicted as pink rods) upon contraction generate external forces $\gv{f}$ and external torques $\gv{Qc}$ on the arm centerline $\gv{x}(s,t)$ \cite{chang2023energy}.
    The longitudinal muscles of radius $r_{m}(s,t):=0.25 r(s,t)$ are attached at $\gv{x}_{m}(s,t) = \gv{x}(s,t) \pm 0.625 r(s,t) \gv{d}_{1}(s,t)$. 
    All 8 arms are actuated by contracting the inside longitudinal muscle, marked in (a), via a traveling wave activation profile of time-period $T$ described in \cref{eq:octopus_actuation}. 
    The octopus head is modeled as a rigid sphere with diameter $D=0.12L$ and center initialized at $(3.15L, 1.25L, 1.25L)$.
    The eight arms and the head of the octopus are connected via the octopus body, which is modeled as a soft tapered Cosserat rod.
    The fluid domain is of size $(5L, 2.5L, 2.5L)$ with the spatial resolution of the domain set as $392 \times 196 \times 196$ throughout.
    We consider two different flow regimes in this study, specifically regimes in which juvenile and adult octopus are observed.
    Key non-dimensional parameters in this benchmark are $t/T$, $\rho=1.02$, actuation Reynolds number $Re_{a} := LD/(T\nu_f)$ (set as 48 for juvenile and 4800 for adult), $K_{b}$ (set as 8.3 for juvenile and 0.083 for adult), and the swimming Reynolds number $Re_{sw} := v_{f} D/ \nu_f$, which emerges as $\order{10}$ for juvenile and $\order{10^2}$ for adult octopus, based on the observed swimming speeds shown in (f).
    Other computational parameters are $\CFL=0.1$, \(\alpha = -2 \times 10^4\) and \(\beta=-10\).
    Snapshots of the volume rendered vorticity magnitude field (white and blue indicates lower and higher vorticity, respectively) around the juvenile octopus for (b) power stroke at $t/T \approx 2.4$ and (c) recovery stroke at $t/T \approx 2.9$. 
    Snapshots of the wake generated by the adult octopus, with the volume rendered vorticity magnitude field for (d) power stroke at $t/T \approx 4.6$ and (e) recovery stroke at $t/T \approx 5.1$.    
    Temporal variation of the octopus head's swimming (f) velocity and (g) position in the adult and juvenile flow regimes.
    Further details can be found under the \texttt{examples/3d_examples/OctopusSwimming} folder of \cite{sopht2023}.
    }
    \label{fig:octopus_swimming}
\end{figure}

Here we demonstrate the versatility of our proposed algorithm in incorporating features of bio-inspired, soft robotic locomotion, namely endogenous biomechanical muscle models, muscular actuations, and soft body assembly, via the case of a self-propelling octopus.
Muscular hydrostats, such as octopus arms, 
are unique in that they lack bones altogether, relying solely on intricate weavings of muscle fibers within their architecture. 
These structures confer unparalleled coordination and reconfiguration abilities,
endowing these animals with remarkable 
manipulation skills. 
These distinctive features have drawn the attention of biologists and soft roboticists to model and characterize the octopus' underwater manipulation and locomotion strategies \cite{kazakidi2015cfd,renda2018unified}. 
However, these studies employ simplistic drag models or neglect elasticity and the arm's internal structure, and thus fall short of resolving the two-way coupling with the flow environment, as well as the effect of the  muscular structure, on the resultant dynamics.

Recently, bio-realistic models based on Cosserat rods have been developed to replicate the heterogeneous organization and the bio-mechanics of the octopus arms \cite{tekinalp2023topology}. 
In particular, these models capture the role of the muscular architecture in
translating non-linear one-dimensional muscle contraction into complex three-dimensional motions, as observed in actual octopuses.
In this study, we represent each octopus arm as a Cosserat rod, and actuate it using a virtual muscle model \cite{chang2023energy} based on the understanding developed in \cite{tekinalp2023topology}.
In the virtual muscle model, the contraction of virtual muscles generates forces $\gv{f}$ and torques $\gv{Qc}$ on the arm center-line \cite{chang2023energy}.
These contractile forces are modeled as
\begin{equation}
f_{m} := a \sigma_{\max} A_{m} h(l_{m}) 
\label{eq:virtual_muscle}
\end{equation}
\noindent
where $a$ denotes the muscle activation level, $\sigma_\max:=10 E$ represents the maximum muscle stress, $E$ is Young's modulus of the arm, $A_{m}:=\pi r_{m}^2$ denotes the muscle cross-sectional area, and $h(l_{m}) \in [0, 1]$ is the normalized force-length relationship obtained by fitting experimental octopus data \cite{kier2002fast}, and $l_{m}$ is the length of the muscle \cite{chang2023energy}. 
This framework allows us to integrate experimentally obtained biomechanical data through the term $h(l_{m})$, while investigating different neuro-muscular activations via the term $a$. 
Next, in order to generate the swimming stroke motion, we use a travelling wave activation profile similar to the one presented in \citet{kazakidi2015cfd}.
\begin{equation}
a(s,t)= a_{\max} \left(\frac{\cos \left(\pi s\right)+1}{2} \right)^4 \sin^{2} \left(2 \pi w \frac{s}{L} - \pi \frac{t}{T} \right)
\label{eq:octopus_actuation}
\end{equation}
\noindent
where $a_{max}=0.2$ is the maximum activation magnitude, $T$ is the period, and $w=0.05$ is the wave number \cite{kazakidi2015cfd}.  
In all our simulations, the octopus starts from a rest configuration and the muscle activation $a_{max}$ is gradually increased from zero to the designated value during the first period. 

In the following, we describe the physical setup of the octopus swimming experiment, as shown in \cref{fig:octopus_swimming}a. 
This computational experiment involves an eight armed octopus immersed in a three-dimensional unbounded viscous flow. Each soft arm is modeled as a tapered Cosserat rod of length $L$. 
The arms and the rigid spherical head are connected to the elastic body through spring-damper boundary conditions. The arms are actuated by contracting their (virtual) longitudinal muscles, following the activation profile specified in \cref{eq:octopus_actuation}.
Additional details about the simulations are provided in the caption of \cref{fig:octopus_swimming}.
Two octopuses are considered in the experiment: an adult and a juvenile. For the adult octopus, the key non-dimensional parameters $K_b$, actuation Reynolds number $Re_{a} := LD/(T\nu_f)$, $AR$, and the taper ratio are based on measurements from experimental studies~\cite{chang2021controlling, tramacere2014structure, kier2007arrangement}. 
The juvenile octopus is a geometrically scaled down version of the adult octopus, roughly $10 \times$ smaller in size \cite{wells1970observations}. 
We then assume that the juvenile octopus has the same activation function $a(s,t)$, $E$, $AR$, and taper ratio as that of the adult, which results in the adult and the juvenile octopus operating in different flow regimes, characterized by distinct values of the arm stiffness $K_{b}$ and the actuation Reynolds number $Re_{a}$.

We begin with visualizing the flow generated by the adult and juvenile octopuses while swimming across the flow domain.
\Cref{fig:octopus_swimming}b presents the power stroke of the juvenile octopus swimming, (visualized via the volume-rendered vorticity magnitude field), where the eight arms generate an octuplet of vortex rings, which are then shed during the recovery stroke shown in \cref{fig:octopus_swimming}c, resulting in forward propulsion of the octopus.
For the adult case instead, an increase in the actuation Reynolds number $Re_{a}$ causes the vortex rings to break down into a jet, as seen from the power and recovery strokes shown in \cref{fig:octopus_swimming}d and e, respectively.
Next, we present the temporal variation of the octopus' swimming velocity and its head's position, in both adult and juvenile flow regimes, in \cref{fig:octopus_swimming}f and g. 
The non-dimensional swimming speed of the juvenile octopus is found to be larger than that of the adult, which is consistent with previous observations of octopuses of different sizes \cite{huffard2006locomotion}.
Overall, this demonstration provides valuable insights into the swimming behavior of octopuses and can be leveraged in the future to develop efficient soft robotic systems inspired by the locomotion strategies of these animals.

\section{Conclusion}\label{sec:conc}

In summary, we have presented a framework that integrates vortex methods and Cosserat rod theory to capture the two-way flow-structure interaction between multiple heterogeneous, slender soft bodies immersed in a viscous fluid. 
Our approach employs the penalty immersed boundary method to couple the fluid and the immersed soft structure, while accurately accounting for the mechanics of soft bodies via Cosserat rod theory.
The algorithm we have developed is straightforward and concise, while still being accurate and robust, as demonstrated by rigorous benchmarking and convergence analysis against a wide range of experimental and numerical tests.
Through various multifaceted illustrations (which themselves may serve as detailed benchmarks for future studies), we further demonstrate our solver’s versatility, applicability and robustness across multiphysics scenarios, boundary conditions, constitutive and actuation models. 
Particularly, the accurate resolution of a wide range of physics encompassing muscle actuation, multi-body connection, self-propulsion, streaming and magnetism in a scalable fashion, demonstrates the utility of our method in a wide range of applications, from bio-locomotion and soft robotics to microfluidics. 
While our algorithm has shown to scale effectively across distributed computing systems, the use of convenient grid-based operators for solving the flow equations renders the solver portable, in a scalable fashion, to advanced parallel architectures such as GPUs. 
However, beyond porting the algorithm to GPUs, opportunities for algorithmic improvement also exist.
First, instead of the current mollification approach, modified higher-order stencils can be utilized to resolve the physics near the interface \cite{xu2006immersed, jiang2018sharp,gabbard2023high} to improve grid convergence now limited between first and second order. 
Second, an unconditionally stable temporal integration algorithm can be applied to solve for the diffusion step~\cite{kolomenskiy2009fourier} to achieve faster time-to-solutions in viscous flow regimes. 
Finally, a time step restriction based on the shear wave speed inside the elastic solid can be bypassed through the use of an implicit time stepper \cite{till2017elastic} or a local time-stepping technique \cite{rossinelli2015mrag}.
All the above directions are avenues of future work.

\section{CRediT authorship contribution statement}
\textbf{Arman Tekinalp:} Conceptualization, Data curation, Formal Analysis, Investigation, Methodology, Software, Validation, Visualization, Writing – original draft, Writing – review \& editing.
\textbf{Yashraj Bhosale:} Conceptualization, Data curation, Formal Analysis, Investigation, Methodology, Software, Validation, Visualization, Writing – original draft, Writing – review \& editing.
\textbf{Songyuan Cui:} Conceptualization, Data curation, Formal Analysis, Investigation, Methodology, Software, Validation, Visualization, Writing – original draft, Writing – review \& editing.
\textbf{Fan Kiat Chan:} Conceptualization, Data curation, Formal Analysis, Investigation, Methodology, Software, Validation, Visualization, Writing – original draft, Writing – review \& editing.
\textbf{Mattia Gazzola:}Conceptualization, Data curation, Formal Analysis, Funding acquisition, Investigation, Methodology, Project administration, Resources, Software, Supervision, Validation, Visualization, Writing – original draft, Writing – review \& editing.
\section{Declaration of competing interest}

The authors declare that they have no known competing financial interests or personal relationships that could have appeared to influence the work reported in this paper.

\section{Acknowledgements}

This study was jointly funded by  ONR MURI N00014-19-1-2373 (M.G.), ONR N00014-22-1-2569 (M.G.), NSF EFRI C3 SoRo \#1830881 (M.G.), NSF CAREER \#1846752 (M.G.), and with computational support provided by the Bridges2 supercomputer at the Pittsburgh Supercomputing Center through allocation TG-MCB190004 from the Extreme Science and Engineering Discovery Environment (XSEDE; NSF grant ACI-1548562).
\appendix

\section{Validation of flow past a two-dimensional elastic rod under gravity}\label{app:bmk_flow_past_2d_rod}

\begin{figure}[!ht]
    \centering
    \includegraphics[width=0.9\textwidth]{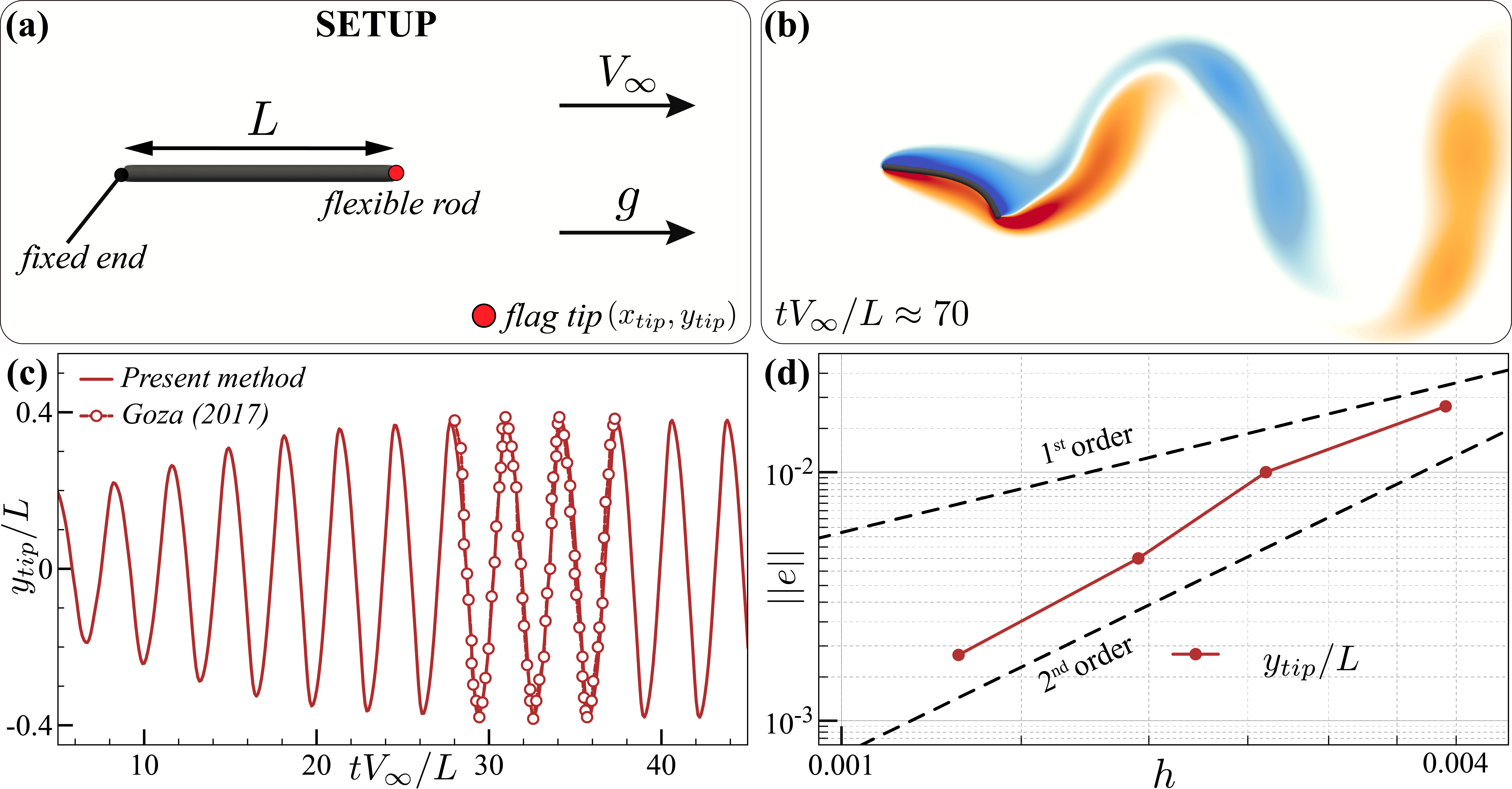}
    \caption{
    Flow past a two-dimensional elastic rod under gravity. (a) Setup. The 2D elastic rod with length $L$, \(\AR = 50\) and mass per unit length $m_e$, is initialized inside a 2D flow domain of size $(6L, 3L)$, with one end fixed at $(L, 1.503L)$ and centerline at $y = 1.503L$. The rod is immersed in a viscous fluid, under a gravity field \(g \hat{\gv{i}}\) and a background free-stream velocity \(V_{\infty} \hat{\gv{i}}\). The key non-dimensional parameters in this benchmark are convective time unit $t V_{\infty} / L$, \(\Rey:= V_{\infty} L / \nu_f = 200\),  \(K_b = 0.0015\), non-dimensional mass ratio \(M_e = m_e / \rho_f L = 1.5\) and Froude number \(Fr:= g L / V_{\infty}^2 = 0.5\). Other computational parameters are \(\CFL = 0.1\), \(\alpha = -8 \times 10^4\) and \(\beta = 30\). (b) Snapshot of the vorticity field (orange and blue indicate positive and negative vorticity respectively) for \(t V_{\infty} L \approx 70\), depicting the rod flapping motion and vortex shedding. (c) Comparison of temporal variation of rod tip's vertical displacement against previous studies \cite{goza2017strongly}. (d) Grid convergence. Error norm of rod tip's vertical displacement amplitude plotted against grid spacing $h$. 
    Further details can be found under \texttt{examples/2d_examples/FlowPastRodCase} folder of \cite{sopht2023}.
    }
    \label{fig:2d_flapping_flag}
\end{figure}

We test our method for the ability to capture the dynamics of two-dimensional elastic structures immersed in viscous fluids. 
To do so we adopt the benchmark of flow past a two-dimensional elastic rod suspended under gravity, commonly used for validation in flag flapping studies \cite{huang2007simulation,goza2017strongly}. 
\Cref{fig:2d_flapping_flag}a presents the initial setup---a two-dimensional elastic rod of length $L$ is immersed in a constant, unbounded, background free stream of velocity \(V_{\infty} \hat{\gv{i}}\) and under the gravitational field \(g \hat{\gv{i}}\). 
The rod is clamped at the upstream end, allowing the rod to flap freely in response to the surrounding flow and gravity. 
Computational setup details can be found in \cref{fig:2d_flapping_flag}.

We simulate the system long enough 
to observe von K\'{a}rm\'{a}n vortices, and eventually the system reaches a dynamically quasi-steady, periodic state. 
This state is visualized via the vorticity contours at a particular time instance, as shown in \cref{fig:2d_flapping_flag}b. 
Next, to validate our method, we track the vertical coordinate of the rod's free tip and compare it against the periodic tip response from previous studies \cite{goza2017strongly}. 
As seen from \cref{fig:2d_flapping_flag}c, our results show close quantitative agreement with the previous study \cite{goza2017strongly}.

We then present the grid convergence for this benchmark, by tracking the amplitude of the rod tip's cross-stream oscillations (\cref{fig:2d_flapping_flag}c), and computing the error norms with respect to the best resolved case. 
We fix $\CFL = 0.1$ and vary the spatial resolution between $256 \times 128$ and $1024 \times 512$ (with $2048 \times 1024$ as the best resolved case). 
As seen from \cref{fig:2d_flapping_flag}d, our method presents grid convergence between first and second order (least squares fit of 2.1), consistent with the spatial discretization of the operators.

\section{Geometrical details of anguilliform swimmers}\label{sec:fish_geometry}

Here we describe the geometry used for the anguilliform swimmers shown in \cref{sec:bmk_fish} of the main text.
The swimmer is 
modeled as an elastic rod with varying cross-section dependent on centerline coordinate $s$.
The two-dimensional swimmer cross-section is characterized using the half width $w(s)$, while the three-dimensional swimmer cross-section is an ellipse with half width $w(s)$ and half height $h(s)$.
Following \citet{gazzola2011simulations}, the half width $w(s)$ is defined as
\begin{equation}
w(s)= \begin{cases}\sqrt{2 w_h s-s^2} & 0 \leqslant s<s_b \\ w_h-\left(w_h-w_t\right)\left(\frac{s-s_b}{s_t-s_b}\right)^2 & s_b \leqslant s<s_t \\ w_t \frac{L-s}{L-s_t} & s_t \leqslant s \leqslant L\end{cases}
\end{equation}
where $L$ is the body length, $w_h = s_b = 0.04 L$, $s_t = 0.95 L$ and $w_t = 0.01 L$.
In \citet{gazzola2011simulations}, the thickness reduction from head to tail ($s_b \leqslant s<s_t$) is linear instead of quadratic for the 2D case, therefore we implemented the same modification here.
For 3D swimmers, we define the half height $h(s)$ following \citet{gazzola2011simulations}
\begin{equation}
h(s)=b \sqrt{1-\left(\frac{s-a}{a}\right)^2}
\end{equation}
where $a = 0.51 L$ and $b = 0.08 L$.

\section{Muscular actuation of anguilliform swimmer}\label{sec:fish_actuation}

We start by creating a virtual rod that is identical to the elastic swimmer, with the former following precisely the kinematics \cref{eq:carling_kinetics} suggested by \citet{Carling:1998}.
Here, we update the directors of the virtual rod under the zero-shear assumption $\left(\gv{d}_{3} \equiv \gv{t}\right)$. 
Next, based on the updated positions and directors we compute the corresponding internal torques $\bar{\gv{\tau}}_v$ within the virtual rod via \cref{eqn:angular_momentum}.
The internal torques of the virtual rod are in turn translated into muscular torques $\gv{\tau}_{m}$ acting upon the elastic swimmer via frame transformation

\begin{equation}
\gv{\tau}_m(s,t) = \gv{Q}_v^T(s,t) \bar{\gv{\tau}}_v(s,t)
\end{equation}

\noindent
where $\gv{Q}^T_{v}$ is the transpose of the virtual rod directors. 
Finally, since the elastic swimmer is free to shear (and therefore may deform slightly different from the virtual rod), local correcting forces are computed, proportionally 
to the deviation between the elastic fish and the virtual rod node positions 

\begin{equation}
    f_{m}(s,t) = k \left( y_{v}(s,t) - y_{e}(s,t) \right)
\end{equation}

\noindent
where $k:=EA$ is the spring constant, $y_{v}$ and $y_{e}$ are lateral displacements of the virtual rod and the elastic fish, respectively. 

\section{Two-dimensional anguilliform swimmer validation}\label{sec:2d_fish_validation_section}

\begin{figure}[!h]
    \centering
\includegraphics[width=\textwidth]{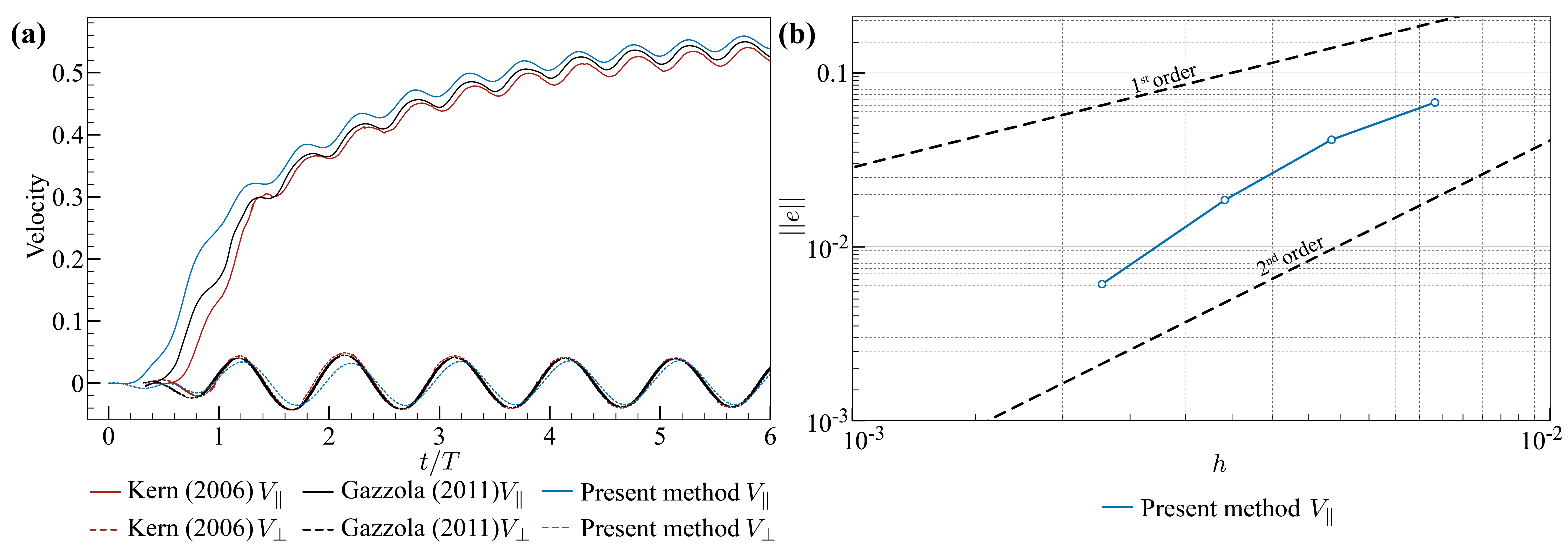}
    \vspace{-0.75cm}
    \caption{
    Self-propelled elastic anguilliform swimmers in two dimensions.
    (a) Validation. Comparison of temporal variation of the elastic swimmer's velocity components ($V_{\parallel}$ and $V_{\perp}$) against previous studies \cite{gazzola2011simulations,Kern:2006} for 2D swimmer.
    (b) Grid convergence. Error norms of $V_{\parallel}$ plotted against grid spacing $h$ for 2D swimmer.
    }  \label{fig:2d_validation_fish_swimming}
\end{figure}

Here we present the validation and convergence analysis for the two-dimensional anguilliform swimmer shown in \cref{sec:bmk_fish}.
For validation, we track the temporal variation of the velocity components \(V_{\parallel} \text{ and } V_{\perp}\) of the 2D swimmer and compare them against previous studies \cite{gazzola2011simulations,Kern:2006}.
As seen from \cref{fig:2d_validation_fish_swimming}a, our results show close agreement with the benchmarks in two dimensions. 
We next present the grid convergence for this case by tracking $V_{\parallel}$ and $V_{\perp}$ and computing the error norms against the best resolved case. 
For two-dimensional swimming, with $\CFL = 0.1$, we vary the spatial resolution  between  $1024 \times 204$ and $2048 \times 409$ (with $3072 \times 614$ being the best resolved case).
As shown in \cref{fig:2d_validation_fish_swimming}b, our technique presents close to second order convergence (least squares fit 
of 1.82 for $V_{\perp}$ and 
2.2 for $V_{\parallel}$, respectively).

\section{Algorithm mapping on distributed computing architectures for large-scale simulations}\label{app:mpi-extension}
In this section, we provide additional details surrounding the mapping of our algorithm onto distributed computing architectures.
In the following, we describe how the flow and immersed body domains are decomposed and communicated across different computing processes.

\subsection{Mapping flow domain}\label{app:flow-domain-decomp}
In the employed numerical method (\cref{alg}), the fluid domain, and hence the flow fields, are defined on structured or uniform Cartesian meshes.
This in turn results in rectangular (2D) or cuboidal (3D) physical and computational domains.

\subsubsection{Domain decomposition}
\begin{figure}[h!]
	\centering
	\includegraphics[width=\linewidth]{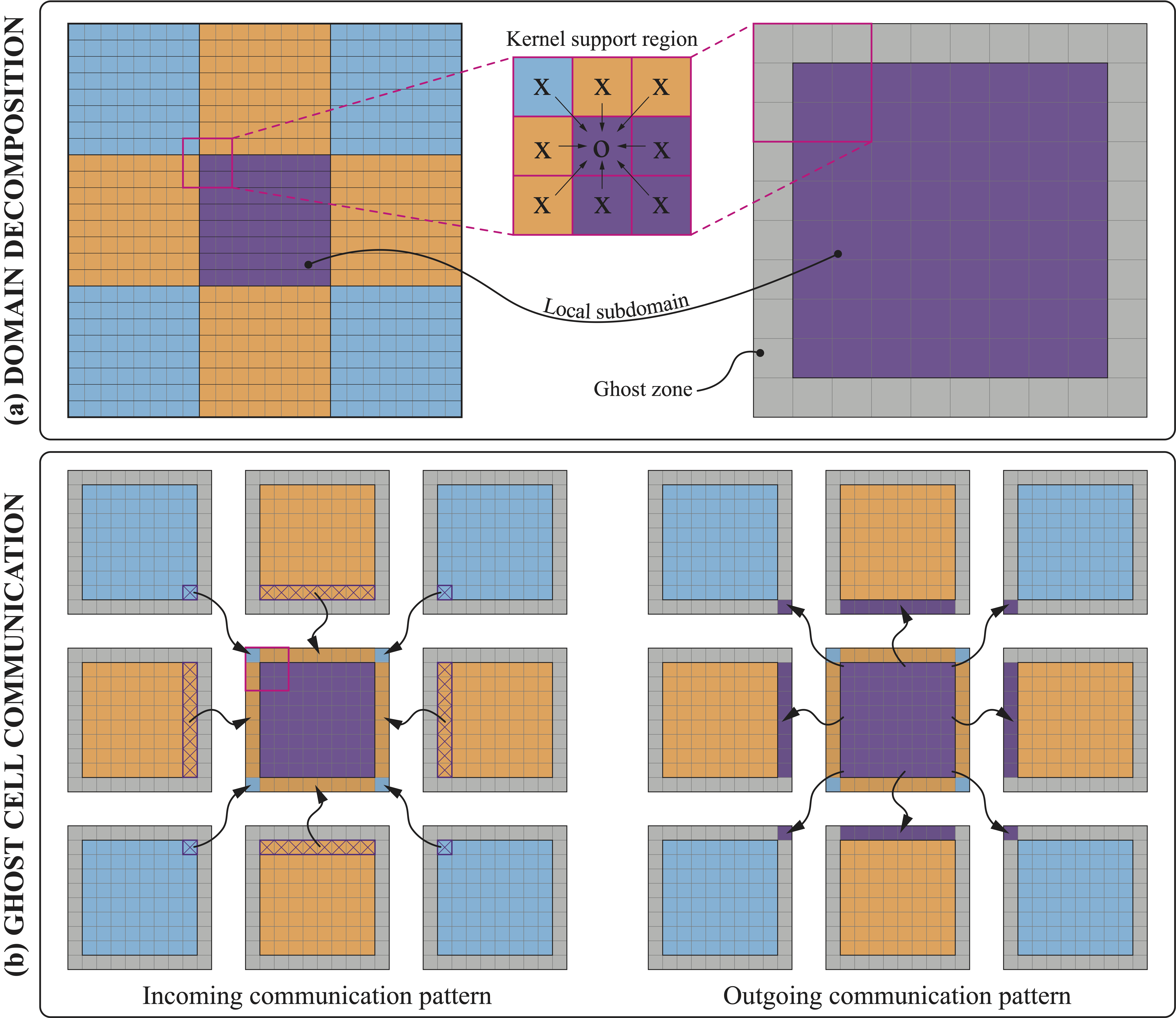}
	\caption{
		(a) Domain decomposition for a representative 2D problem. We focus our example here on the purple central subdomain and use a nine-point stencil to motivate the need for communication across subdomains residing on different processes. We introduce ghost cells for storing data copied from neighboring subdomains in order to ensure the correct computation of differential operators.
		(b) Ghost cell communication pattern for complete data exchange: receiving (left)incoming  and sending (right) outgoing data.
	}
	\label{fig:sopht-mpi-eulerian-comm}
\end{figure}

In order to ensure balanced workloads between processors, we employ the Cartesian topology available via the Message Passing Interface (MPI) \cite{Dalcin:2021,Dalcin:2011,Dalcin:2005,Dalcin:2008} for decomposing the fluid domain into equally-sized subdomains that are mapped onto different processors (\cref{fig:sopht-mpi-eulerian-comm}a).
Given the assigned subdomain, each processor then performs differential operations with mesh-based kernels (i.e. numerical differential stencils).
These local operations typically result in computational costs that scale linearly with the number of grid points.
However, in order to obtain the correct solution on the boundaries of these subdomains, communication with the adjacent subdomain becomes necessary to ensure the local operations are carried out with the correct boundary data.

\subsubsection{Communication}
In order to exchange the boundary data between neighboring subdomains, we extend the subdomain by $n_s$ mesh points (grey regions of subdomains in \cref{fig:sopht-mpi-eulerian-comm}), where $n_s$ is chosen based on the maximum kernel support required by the numerical stencils employed in \cref{alg}.
For example, a second-order central finite difference stencil requires one adjacent mesh point for its kernel operation and thus has a kernel support of $n_s = 1$.
This extended layer of mesh points is commonly known as ghost/halo points (\cref{fig:sopht-mpi-eulerian-comm}a), and stores updated copies of the data from the corresponding neighboring subdomains for consistent kernel operation on the boundaries.

As an illustrative example, we show in \cref{fig:sopht-mpi-eulerian-comm} the communication pattern involved for a 2D problem decomposed and distributed equally into 9 different processors.
We start by drawing our attention to the communication patterns involved for the representative interior (purple) subdomain in \cref{fig:sopht-mpi-eulerian-comm}, and note that the communication pattern generically holds for other subdomains as well.
The orange and blue regions indicate the subdomains directly and diagonally adjacent to the interior subdomain, respectively.
We then consider a representative nine-point stencil kernel (pink box in \cref{fig:sopht-mpi-eulerian-comm}a) that operates on the structured mesh.
As can be seen, in the extreme case of kernel operation at corner cells of our focal subdomain, the stencil involves information from a total of three neighboring processes.
This means that for the operator to be computed correctly on that corner cell, communication needs to happen between the purple subdomain, two directly (orange) and one diagonally (blue) adjacent subdomains.
\Cref{fig:sopht-mpi-eulerian-comm}b illustrates the data flow for updating the ghost region that ensures correct computation of kernels with $n_s > 0$ across all subdomains.

\subsection{Mapping immersed body domain}\label{subsec:body-domain-decomp}
The immersed structures in the employed numerical method (\cref{alg}) are defined on unstructured meshes.
In order to resolve the structural dynamics of the slender immersed bodies, we leverage \texttt{pyelastica} \cite{pyelastica2023}, a simple and convenient library for simulating assemblies of slender, one-dimensional structures using Cosserat Rod theory.
However, the package currently does not support distributed computation.
While it is reasonable to assign the Cosserat rod simulation on a single master process, information from the updated structure will eventually be needed in corresponding flow subdomains during the flow--structure interaction step in \cref{alg}.
For the interaction to take place correctly, our algorithm requires the position, velocity and forces defined on the unstructured mesh 
native to the immersed body.
In the following, we describe how the immersed structure is distributed to and retrieved from different processes throughout the interaction step in our method.

\begin{figure}[h!]
	\centering
	\includegraphics[width=\linewidth]{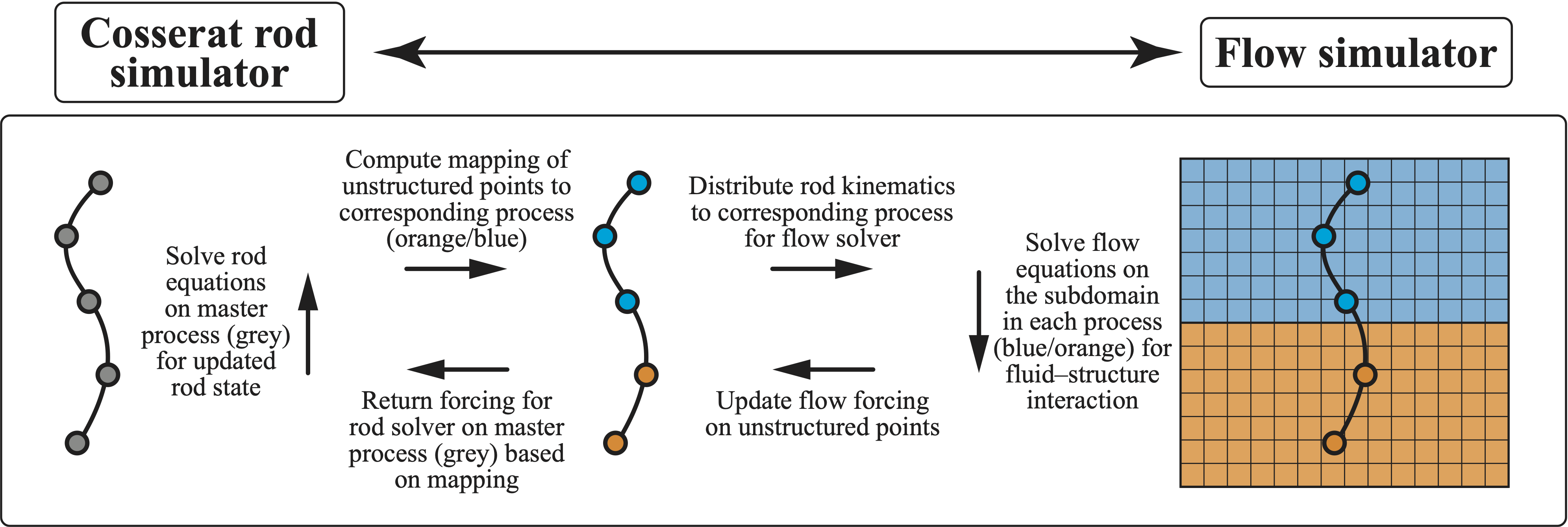}
	\caption{Communication pattern involved for connecting structured and unstructured grids, thus allowing extension of flow simulator to interact with external structural dynamics packages such as \texttt{pyelastica} \cite{pyelastica2023} for Cosserat rod simulations.}
	\label{fig:sopht-mpi-lagrangian-comm}
\end{figure}

\subsubsection{Domain decomposition}
Given the dynamic, unstructured nature of the moving immersed body, the master process will need to decompose the body's discretization mesh based on the position of the body's grid points relative to the fluid subdomain, and remap them to the corresponding fluid subdomain at every time step.
This is done in a straightforward fashion based on the Cartesian topology established by MPI \cite{Dalcin:2005,Dalcin:2008,Dalcin:2011,Dalcin:2019} for domain decomposition of the structured fluid domain (\cref{app:flow-domain-decomp}).

\subsubsection{Communication}
Once the decomposition map is computed, relevant fields residing on the unstructured mesh can be communicated to and from their corresponding subdomain for consistent resolution of flow--structure interaction.
We summarize the communication pattern for a simple unstructured body mesh in \cref{fig:sopht-mpi-lagrangian-comm}.
As described above, the master process first identifies the corresponding flow subdomain to which every single point of the body grid resides in.
This information is then stored in a mapping containing the address of destination processes for each body grid point.
Based on this mapping, the mesh points are communicated to the corresponding flow subdomain, followed by the flow--structure interaction step of \cref{alg}.
Once the interaction is complete, relevant updated fields such as forcing on the structure are returned from the subdomains back to the master process accordingly based on the mapping established earlier.
With the forcing information updated on the master process, \texttt{pyelastica} \cite{pyelastica2023} then performs temporal integration to advance the rod to the next time step, resulting in an updated position and velocity of the rod's unstructured mesh.
This process is then repeated at every time step, thus ensuring a continuous and consistent workflow for the fluid--structure interaction.

\section{Validation for flow past a rigid sphere}\label{app:flow_past_sphere}

\begin{figure}[!ht]
    \centering
    \includegraphics[width=\textwidth]{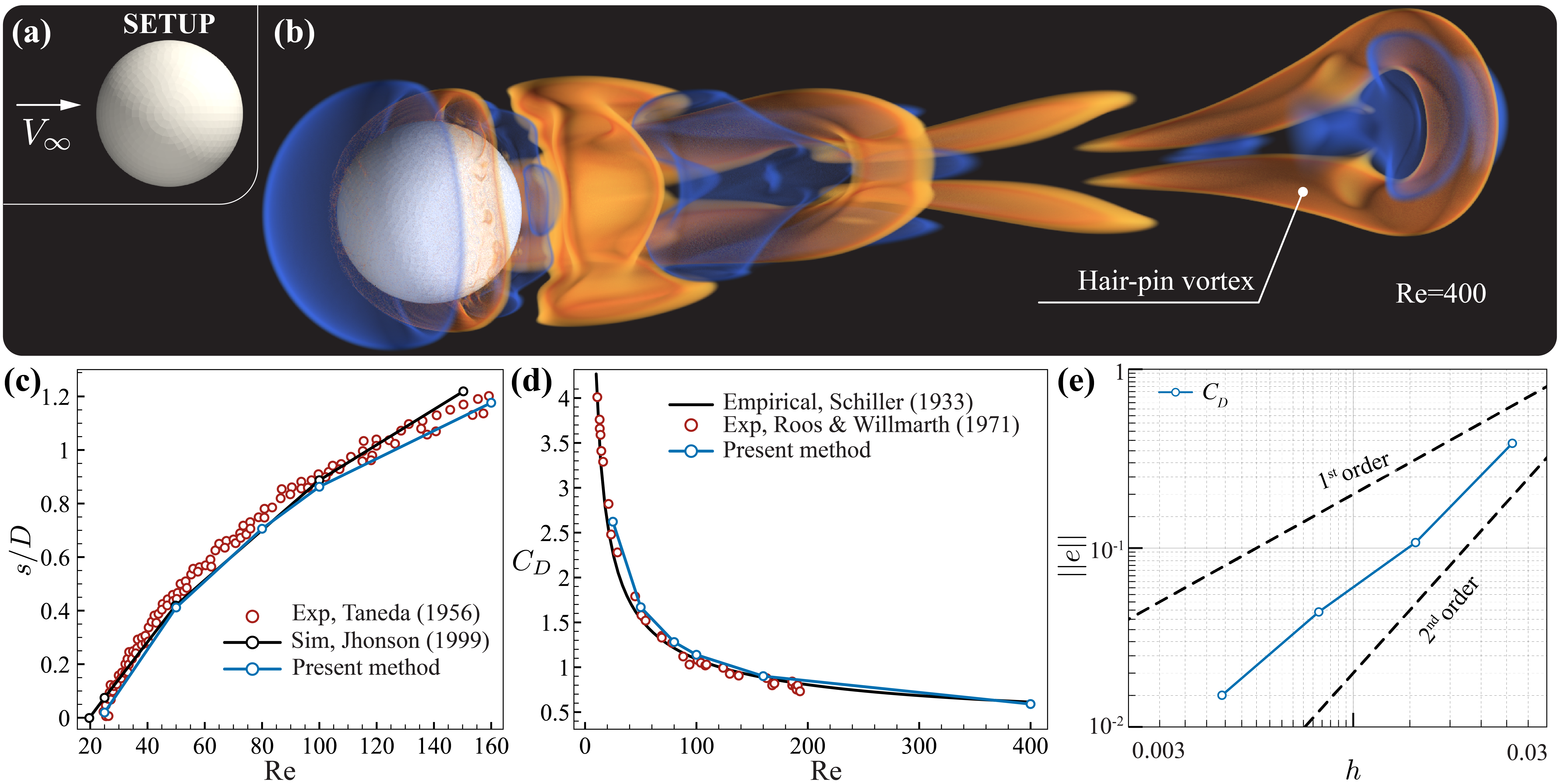}
    \caption{
    Flow past a rigid sphere.
    (a) Setup. The rigid sphere with diameter $D$ is initialized inside a 3D flow domain of size $(5D, 2.5D, 2.5D)$, and with its center fixed at $(1.25D, 1.25D, 1.25D)$.
    The sphere is immersed in a viscous flow with a free-stream velocity $V_{\infty} \hat{\gv{i}}$.
    The key non-dimensional parameters in this benchmark is $Re := V_{\infty} D / \nu_f$.
    Other computational parameters are CFL=0.1, $\alpha=-1.5 \times 10^{5}$ and $\beta=-87.5$.
    (b) Snapshot of flow past sphere at $Re=400$ with volumetric rendering of Q-criterion (orange and blue indicate vorticity and strain dominated areas, respectively).
    To capture the far-field flow features, we run this particular simulation with a flow domain of size $(10D, 4.2D, 4.2D)$, spatial resolution $384 \times 160 \times 160$, and sphere center fixed at $(2.15D, 2.1D, 2.1D)$.
    Formation of hair pin vortex in the wake is captured, consistent with previous studies \cite{Johnson:1999}.
    (c) Comparison of non-dimensional attached eddy length $(s/D)$ vs $Re$ against experiments (red circles) \cite{Taneda:1956} and previous simulations (black line with circle markers) \cite{Johnson:1999}.
    (d) Comparison of mean drag coefficient $(C_{D})$ vs $Re$ against empirical (black line) \cite{Schiller:1933} and experiments (red circles) \cite{Roos:1971}.
    (e) Convergence: Error norms of sphere $C_{D}$ for $Re=100$ plotted against grid spacing $h$.     
    Further details can be found under \texttt{examples/3d_examples/FlowPastSphereCase} folder of \cite{sopht2023}.
    } \label{fig:3d_valid_flow_past_sphere}
\end{figure}

Here we test our algorithm for its ability to capture flow--structure interaction physics for immersed rigid bodies, by considering the classical case of flow past a fixed rigid sphere.
\Cref{fig:3d_valid_flow_past_sphere}a presents the initial setup---a fixed rigid sphere of diameter $D$ is immersed in a constant, unbounded, background free stream of velocity \(V_{\infty} \hat{\gv{i}}\) (details in \cref{fig:3d_valid_flow_past_sphere} caption).
We simulate the system long enough until it reaches either a steady state (observed for $\Rey < 200$) or a quasi-steady periodic state (observed for $200 < \Rey < 1000$).
For low $\Rey$ ($< 200$), a steady recirculation zone (eddy) attaches to the downstream end of the sphere, while for moderate $\Rey$ ($200 < \Rey < 1000$), periodic shedding of vortices is observed in the wake of the sphere.
The periodic vortex shedding state is visualized for a particular $\Rey$ via the volume rendered Q-criterion in \cref{fig:3d_valid_flow_past_sphere}b.
As seen from \cref{fig:3d_valid_flow_past_sphere}b, hair-pin vortices are observed in the wake of the sphere, consistent with previous studies \cite{Johnson:1999}. 
Next, for validation, we vary $\Rey$ and track two commonly used non-dimensional diagnostics: the length of the attached recirculation eddy $s/ D$ (only for $\Rey < 200$) and the mean drag coefficient $C_D$ for the sphere. 
\Cref{fig:3d_valid_flow_past_sphere}c-d present the variations of recirculation eddy lengths and mean drag coefficients with $\Rey$, with comparison against previous studies \cite{Johnson:1999,Taneda:1956,Roos:1971,Schiller:1933}. 
As seen from \cref{fig:3d_valid_flow_past_sphere}c-d, our results show close agreement with previous studies.

Next, we present the grid convergence for this case, by tracking the mean drag coefficient $C_D$ for $\Rey = 100$, and computing the error norms against the best resolved case.
With $\CFL = 0.1$, we vary the spatial resolution between $32 \times 16 \times 16$ and $256 \times 128 \times 128$ (with $512 \times 256 \times 256$ as the best resolved case).
As seen from \cref{fig:3d_valid_flow_past_sphere}e, our method exhibits spatial convergence between first and second order (least square fit of 1.53), consistent with the numerical discretization of our solver.

\section{Animations of simulated cases}\label{app:videos}
We have included animated videos of all cases simulated in the paper,
which can be accessed through the \href{https://drive.google.com/drive/folders/1QuYzb0YtS-s7NBHhEeWTJI1EgQ0fCxvt?usp=sharing}{following link}.

\section{Software}\label{app:software}
The implementation of the algorithm in this study and corresponding cases can be found \href{https://github.com/SophT-Team/SophT}{here}. For the MPI based implementation the reader is referred to this \href{https://github.com/SophT-Team/sopht-mpi}{link}.

\bibliographystyle{elsarticle-num-names}
\bibliography{cfs_lit.bib}





\end{document}